\documentclass[12pt]{article}
\pdfoutput=1
\usepackage{color}
\usepackage{mhchem}
\usepackage{xcolor}
\usepackage{cite}
\usepackage{hyperref}
\hypersetup{colorlinks=true,linkcolor=red,anchorcolor=black,citecolor=green}
\usepackage[toc,page]{appendix}
\usepackage{mathtools}
\usepackage{amsfonts}
\usepackage{bbold}
\usepackage{textcomp}
\usepackage[DIV13]{typearea}
\usepackage{amsmath, amsthm, amssymb, mathtools,empheq,latexsym,dsfont}
\usepackage{bbm}
\usepackage{mathrsfs}
\usepackage{slashed, simplewick}
\usepackage[utf8]{inputenc}
\usepackage{graphicx,placeins}
\usepackage{makeidx}
\usepackage[font=small,labelfont=bf]{caption}
\usepackage{nicefrac}
\usepackage{subfigure}
\usepackage{array, bigdelim,multirow,multicol}
\usepackage[integrals]{wasysym}
\usepackage{fancybox}
\usepackage{bm}
\usepackage{float}
\usepackage{rotating}
\usepackage{colortbl}
\usepackage{booktabs}
\usepackage[top=2cm,textwidth=16.6cm,textheight=22.75cm]{geometry}
\usepackage{doi}
\graphicspath{{immagini/}}

\definecolor{Gray}{gray}{0.92}
\def\dd{\displaystyle}
\newcommand{\nn}{\nonumber}
\newcommand{\be}{\begin{equation}}
\newcommand{\ee}{\end{equation}}
\newcommand{\bea}{\begin{eqnarray}}
\newcommand{\eea}{\end{eqnarray}}

\newcommand{\dmu}{\partial_\mu}
\newcommand{\osmu}{\overline{\sigma}^\mu}

\newcommand{\of}{\overline{f}}
\newcommand{\Kf}{K^f(\varphi)}

\makeatletter
\renewcommand*{\@fnsymbol}[1]{\ensuremath{\ifcase#1\or *\or  \mathsection\or \ddagger\or
\dagger\or \mathparagraph\or \|\or **\or \dagger\dagger
\or \ddagger\ddagger \else\@ctrerr\fi}}
\makeatother

\begin{document}
 \unitlength = 1mm

\setlength{\extrarowheight}{0.2 cm}

\title{
\begin{flushright}
\hfill\mbox{{\small\tt USTC-ICTS/PCFT-20-07}}\\[5mm]
\begin{minipage}{0.2\linewidth}
\normalsize
\end{minipage}
\end{flushright}
 {\Large\bf Testing Moduli and Flavon Dynamics\\[0.3cm]
  with Neutrino Oscillations}\\[1.3cm]}
\date{}

\author{
Gui-Jun~Ding$^{1,2}$
\thanks{E-mail: {\tt dinggj@ustc.edu.cn}}
\ and
Ferruccio~Feruglio$^{3}$
\thanks{E-mail: {\tt feruglio@pd.infn.it}}
 \
\\*[20pt]
\centerline{
\begin{minipage}{\linewidth}
\begin{center}
$^1${\small Peng Huanwu Center for Fundamental Theory, Hefei, Anhui 230026, China} \\[2mm]
$^2${\small
Interdisciplinary Center for Theoretical Study and  Department of Modern Physics,\\
University of Science and Technology of China, Hefei, Anhui 230026, China}\\[2mm]
$^3${\small
 Dipartimento di Fisica e Astronomia `G.~Galilei', Universit\`a di Padova\\
INFN, Sezione di Padova, Via Marzolo~8, I-35131 Padua, Italy}\\
\end{center}
\end{minipage}}
\\[10mm]}
\maketitle
\thispagestyle{empty}

\centerline{\large\bf Abstract}
\begin{quote}
\indent
We study scalar Non-Standard Neutrino Interactions (NSI) induced by moduli or flavon exchange between electrons and neutrinos. In a region with non-vanishing electron number density, they are known to determine a shift of the neutrino mass matrix. We review and extend the relevant formalism, and we update the existing limits on electron and neutrino scalar couplings. We explore the observability of scalar NSI in models of lepton masses based on flavour symmetries. We analyze models where the scalar couplings are constrained either by abelian symmetries or by modular invariance. We highlight regions of the parameter space where observable effects can occur.
\end{quote}

\newpage

\section{Introduction}
In most of the frameworks aiming to a solution of the flavour puzzle, fermion masses are field-dependent quantities
assuming specific values once the vacuum of the theory is selected. Yukawa couplings depends on the
vacuum expectation values (VEVs) of a set of scalar fields $\varphi$, new dynamical degrees of freedom
predicted or postulated in the underlying theory.
For example, in string theory coupling constants are naturally field-dependent objects and the scalar fields $\varphi$ can be moduli, describing shape and size of compactified extra dimensions.
In a bottom-up perspective, the couplings of $\varphi$ to the Standard Model (SM) fermions are often constrained by flavour symmetries and the observed pattern of fermion masses and mixing angles represents the effect of breaking a symmetry group acting in generation space.
The scalar fields $\varphi$, called flavons in this context, have non-trivial transformation properties under the group, acquire non-vanishing VEVs $\varphi_0$ and break the flavour symmetry.
The observed fermion masses are shaped by the flavons VEVs $\varphi_0$, due to the restricted functional
dependence of Yukawa couplings on $\varphi$.
In this class of models, the new scalar degrees of freedom are mandatory,
given the absence of realistic unbroken flavour symmetries \cite{Feruglio:2019ktm}.
In string theories, flavour symmetries can arise from isometries of the compactified space or from selection rules \cite{Kobayashi:2006wq,Nilles:2012cy,Baur:2019kwi}, thus restricting the choice of possible flavour groups and flavon representations.

It would be highly desirable to test this scenario, by directly accessing to VEVs, masses and couplings of the
new scalar sector. The characteristic scale of the flavour symmetry breaking sector is unknown and often
assumed to be very large, to avoid problems with new potential sources of flavour-changing neutral currents.
The typical coupling constants, arising from higher dimensional operators suppressed by the flavour scale,
can be very small, further reducing the prospect of detectability of the new degrees of freedom.

Large scales and small scalar couplings leading to scalar-mediated non-standard neutrino interaction (NSI) can in principle be tested in neutrino oscillations.
Scalar NSI are known to modify neutrino masses
when neutrinos propagate in matter \cite{Sawyer:1998ac}, at variance with NSI mediated by vector particles that
affect the Wolfenstein potential. The framework is analogous to that of mass-varying neutrinos  \cite{Fardon:2003eh,Kaplan:2004dq},
invoked to link neutrino and dark energy densities, whose impact on neutrino oscillations have been analyzed in \cite{Barger:2005mn,GonzalezGarcia:2006vp}.
General NSI, also including scalar interactions, have been studied in ref. \cite{Bergmann:1999rz}.
More recently, scalar NSI have been reconsidered in ref. \cite{Ge:2018uhz} as a possible source
of deviations in neutrino oscillations. Important features have been pointed out in refs. \cite{Smirnov:2019cae}
and \cite{Babu:2019iml}.

Neutrinos inside an infinite region filled by electrons with constant electron number density $n_e^0$, experience a mass shift
\be
\delta m_\nu=-n_e^0 \frac{{\tt Re}({\cal Z}^e){\cal Z}^\nu}{M^2} ~~~,
\ee
where ${\cal Z}^e$ and ${\cal Z}^\nu$ are the couplings of the scalar field to electrons and neutrinos in a two-component spinor notation and $M$ is the mass of the scalar particle. To produce a shift of few meV in a region with an electron number density close to the one in the sun,
an effective coupling ${\tt Re}({\cal Z}^e){\cal Z}^\nu/M^2\approx 10^4$ GeV$^{-2}$ is required.
This is more than eight orders of magnitude larger than the Fermi constant, representing the first big obstacle in our
task. The reason why a very large effective interaction is needed resides in the different energy dependence
between scalar and vector NSI, the former being depleted by an approximate factor $m_\nu/E$ compared to the latter.
An immediate possibility that comes to mind to enhance the effective coupling is to consider a very light scalar mediator.
Here comes the second obstacle, related to the inevitable finite size $L$ of the the region with a non-negligible electron number density.
As pointed out in ref. \cite{Smirnov:2019cae}, when the Compton wavelength $\lambda=\hbar/(M c)$ of the mediator
becomes larger than $L$, the effective coupling constant approaches
${\tt Re}({\cal Z}^e){\cal Z}^\nu L^2 c^2/\hbar^2$. There is no more gain in lowering the scalar mass below $\hbar/(L c)$.
The third obstacle is represented by the formidable limits that current tests of gravity set on the coupling of an ultralight
scalar to electrons. Both tests of the inverse square law (ISL) of gravity and of the equivalence principle (EP) are
very effective in bounding $|{\tt Re}({\cal Z}^e)|$, which, in the region of interest, cannot exceed too much the tiny value $10^{-25}$.
Neutrino interactions to light scalars are less severely bounded, but important limits exist from the well-established
free-streaming property of neutrinos following their decoupling in the early universe.

In the light of the previous discussion, the perspective of detecting scalar NSI through their effect in neutrino oscillations seem very reduced,
even more so if studied in the context of a specific model, where the pattern of couplings is largely dictated by symmetry considerations.
The purpose of the present work is to investigate in detail this possibility, by examining some representative
class of models of lepton masses based on flavour symmetries.
In Section \ref{S2} we shortly review the formalism of scalar NSI, following closely the discussion of ref. \cite{Smirnov:2019cae} and \cite{Babu:2019iml}. We also include new considerations on the existing limits on electron and neutrino couplings
to a very light scalar particle. In Section \ref{S3} we analyze scalar NSI in the context of models of lepton masses based
on flavour symmetries. Here the discussion is completely general and covers the case of theories containing multiple scalars and allowing non-canonical kinetic terms. This considerably extends the existing formalism. In Section \ref{S4} we analyze
models with an abelian flavour symmetry group. We first discuss a toy model, to show the main problems related
to the detectability of a signal, and then we move to an example where observable scalar neutrino NSI are possible.
In Section \ref{S5} we consider models of lepton masses based on modular invariance.
We consider this application particularly relevant, given the opportunity of directly testing the dynamics
of the modulus, the unique symmetry breaking parameter of this class of theories. We derive in full generality the modulus-lepton coupling and we apply the formalism to a case study in Section
\ref{S6}. Finally in Section \ref{S7} we discuss our results, stressing strengths and limitations of our analysis.
\section{Neutrino masses and scalar interactions}
\label{S2}
In view of the very recent developments and for the sake of clarity, we shortly review
in this section the discussion of ref. \cite{Smirnov:2019cae} and \cite{Babu:2019iml}, which is very relevant for our analysis.
We also complement this review with additional considerations on the existing limits on electron and neutrino couplings
to a very light scalar particle.

We consider a set of real scalars $\varphi_\alpha$ interacting with electrons and neutrino, with field dependent
masses $m_{e,\nu}(\varphi_\alpha)$. By expanding  $m_{e,\nu}(\varphi)$ around the minimum $\varphi_\alpha^0$ up to first order in the fluctuations, we have $m_{e,\nu}(\varphi)=m_{e,\nu}+{\cal Z}^{e,\nu}_\alpha\varphi_\alpha+...$
\footnote{We set $m_{e,\nu}=m_{e,\nu}(\varphi^0)$ and, to simplify the notation, we redefine the fluctuation $(\varphi_\alpha-\varphi_\alpha^0)$ as $\varphi_\alpha$.}.
Assuming Majorana neutrinos and adopting the two-component spinor notation, the Lagrangian reads:
\bea
{\cal L}&=&i\sum_{f=e,e^c,\nu}\of~\osmu \dmu f+\frac{1}{2}\partial_\mu \varphi_\alpha\partial^\mu\varphi_\alpha-\frac{1}{2}M_\alpha^2 \varphi_\alpha^2\nn\\
&-& (m_e+{\cal Z}^e_\alpha \varphi_\alpha) e^c e -\frac{1}{2}\nu(m_\nu+{\cal Z}^\nu_\alpha\varphi_\alpha)\nu+h.c.+...\,.
\label{eq:NSI}
\eea
Here $e$ and $e^c$ describe the first generation charged leptons, while $\nu$ is a multiplet in generation space. Similarly, for each $\alpha$, ${\cal Z}^e_\alpha$ is a number, while ${\cal Z}^\nu_\alpha$ is a 3$\times$3 symmetric matrix. In a more general setting, electron and neutrinos have non-canonical kinetic terms, depending on the
fields $\varphi_\alpha$, which induces an additional dependence of the electron and neutrino interaction on $\varphi_\alpha$. After standard field redefinitions, which will be described in the next Section, it is always possible
to put the Lagrangian into the form (\ref{eq:NSI}) given above, which we use as a starting point of our discussion.
The equations of motion of neutrinos and scalars are:
\bea
&&i\sigma^\mu \partial_\mu \bar \nu-(m_\nu+{\cal Z}^\nu_\alpha\varphi_\alpha) \nu=0\nn\\
&&-(\Box+M_\alpha^2)\varphi_\alpha-\left({\cal Z}^e_\alpha e^c e+\frac{1}{2}\nu{\cal Z}^\nu_\alpha \nu+h.c.\right)=0~~~.
\label{eom}
\eea
Assuming a static unpolarized background with negligible neutrino number density, the second equation becomes
\be
(\nabla^2-M_\alpha^2)\varphi_\alpha={\tt Re}({\cal Z}^e_\alpha) n_e(\vec x)~~~,
\ee
solved by
\be
\varphi_\alpha(\vec x)=-{\tt Re}({\cal Z}^e_\alpha)\int d^3 x'
\frac{e^{\dd-M_\alpha\vert\vec x-\vec x'\vert}}{4\pi\vert\vec x-\vec x'\vert}n_e(\vec x')~~~.
\label{solphi}
\ee
By making use of the first equality in eq. (\ref{eom}), as a result of the scalar exchange we get a shift
of the neutrino mass matrix given by:
\be
\delta m_\nu(\vec x)=\sum_\alpha {\cal Z}_\alpha^\nu \varphi_\alpha(\vec x)~~~.
\ee
To understand the qualitative behavior of this solution it is instructive to consider the simple case of a
constant electron number density $n_e^0$, vanishing outside a spherical region of radius $R$ centered
at the origin. Evaluating $\varphi_\alpha$ at $\vec x=0$ we find:
\bea
\varphi_\alpha(0)&=&-\frac{n_e^0}{M_\alpha^2}{\tt Re}({\cal Z}^e_\alpha)F(M_\alpha R)\nn\\
\label{eq:FF_constdensity} F(M_\alpha R)&=&1-e^{\dd -M_\alpha R}(1+M_\alpha R)\approx
\left\{
\begin{array}{cc}
1&M_\alpha\gg 1/R\\
M_\alpha^2 R^2/2& M_\alpha\ll1/R
\end{array}
\right.~~~.
\eea
If the Compton wavelength $\hbar/(M_\alpha c)$ is smaller that $R$, we have the $1/M_\alpha^2$ suppression
expected form a Yukawa potential, while for $\hbar/(M_\alpha c)$ much larger than $R$, the potential due to
the scalar exchange is indistinguishable from the Coulomb one and proportional to $(3R^2-|\vec x|^2)/6$ in the interior of the sphere. This distinction, stressed in ref. \cite{Smirnov:2019cae}, is very important for the application examined here. For fixed values of the coupling constants, the potential cannot be made arbitrarily large by taking tiny scalar masses. Any realistic physical system has a finite size $R$ and when $M_\alpha$ becomes much smaller than $1/R$, the behaviour $1/M_\alpha^2$ is cut off and replaced by $R^2$.
For example in the Sun(Earth) we have $R\approx 6.955\times 10^5(6.378\times 10^3)$ Km, which corresponds to $1/R\approx 2.84\times 10^{-16}(3.09\times 10^{-14})$ eV.
Neutrinos at the center of the above idealized region experience a mass shift
\be
\delta m_\nu(0)=-n_e^0 \sum_\alpha  \frac{{\tt Re}({\cal Z}^e_\alpha)}{M_\alpha^2}F(M_\alpha R) {\cal Z}^\nu_\alpha~~~.
\label{dcenter}
\ee

\begin{figure}[t!]
\centering
\includegraphics[width=0.65\textwidth]{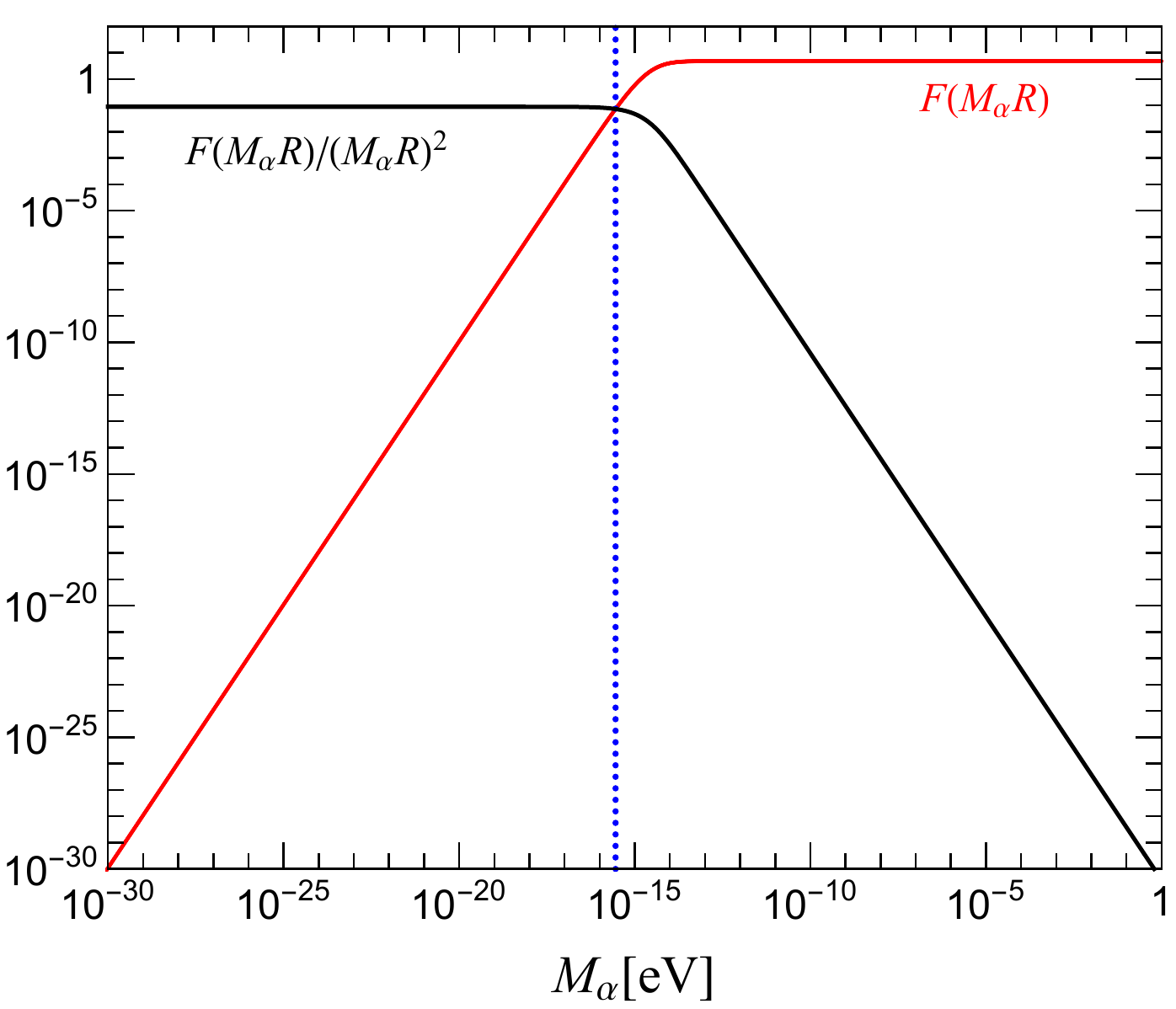}
\caption{Dependence of the factors $F(M_{\alpha}R)$ and $F(M_{\alpha}R)/(M_{\alpha}R)^2$ on the scalar mass $M_{\alpha}$ for the sun, from eq. (\ref{eq:FF}), choosing $n^0_e=10^{11}\mathrm{eV}^3$.
We take $R=6.955\times 10^5$ Km and $n_e(r)$ from ref. \cite{Bahcall:2004pz,Bahcall_web}. The vertical blue dotted line denotes $M_{\alpha}=1/R\simeq2.84\times10^{-16}$ eV.}
\label{fig:Ffactor}
\end{figure}

In our application we are interested in a spherically symmetric, not necessarily constant,
matter distribution: $n_e(\vec x)=n_e(r)$, like the one in the sun or in the Earth. In this case
the solution (\ref{solphi}) can be made more explicit \cite{Smirnov:2019cae}:
\begin{eqnarray}
\varphi_\alpha(r)=-\frac{{\tt Re}({\cal Z}^e_\alpha)}{M_{\alpha}r}\left[e^{-M_{\alpha}r}\int^{r}_0x~ n_e(x)\sinh(M_{\alpha}x)dx+\sinh(M_{\alpha}r)\int^{+\infty}_{r}x~ n_e(x)e^{-M_{\alpha}x}dx\right]\,.
\end{eqnarray}
We can still write the shift of the neutrino mass matrix at the center of the region as in eq. (\ref{dcenter}),
with the factor $n_e^0 F(M_{\alpha}R)$ given by
\begin{equation}
\label{eq:FF}
n_e^0 F(M_\alpha R)=M_\alpha^2R^2\int^{1}_{0}ye^{\dd-(M_{\alpha}Ry)}~n_e(Ry)dy\,,
\end{equation}
where we assumed $n_e(r)$ vanishing for $r>R$. The electron number density $n_e$ is given by $n_e=Y_e\rho/m_p$,
where $\rho$ is the density of the matter, $m_p=938.27~\mathrm{MeV}$ is the proton mass, $Y_e=N_e/(N_n+N_p)$ is the electron fraction (or the number of electron per nucleus), and $Y_{e}\sim0.5$ for neutral matter. Typical values for the matter density are $\rho_{\mathrm{crust}}\simeq 3\mathrm{g/cm^3}$ in the Earth's crust, and $\rho_{\mathrm{sun}}\simeq 150 \mathrm{g/cm^3}$ in the sun core. Consequently the electron number density in the sun (earth) is of order $n^0_e=10^{11}\mathrm{eV}^3$ ($10^{9}\mathrm{eV}^3$).

Throughout this paper we use eq. (\ref{eq:FF}) as a definition of $F(M_\alpha R)$, by choosing as a reference density $n_e^0=10^{11}$ eV$^3$ for the sun. Of course, only the product $n_e^0 F(M_\alpha R)$ has a physical meaning and the choice of $n_e^0$ is purely conventional. We have computed $F(M_\alpha R)$ for different values of  $M_{\alpha}$ in the center of the sun using the electron density distribution from~\cite{Bahcall:2004pz,Bahcall_web}. From fig. \ref{fig:Ffactor} we see that the factor $F(M_{\alpha}R)$ tends to a constant value when $M_{\alpha}\gg 1/R$ and is approximately proportional to $M^2_{\alpha}R^2$ for $M_{\alpha}\ll1/R$. The asymptotic behavior of $F(M_{\alpha}R)$ agrees well with Eq.~\eqref{eq:FF_constdensity} derived under the assumption of constant electron density. From Eq.~\eqref{eq:FF} we see that this asymptotic behavior of $F(M_{\alpha}R)$ should generally hold true for any physical system with spherically symmetric matter density distribution and finite size.

Scalar interaction of electrons and neutrinos are severely constrained. As we will see,
the scalar masses we are interested cover the region from $10^{-4}$ eV to $10^{-22}$ eV. In this range the main
bounds on the relevant coupling come from astrophysics and cosmology and they will be recalled in the next
sub-sections.
\subsection{Limits on electron coupling}
\label{LRF}
A first bound on the electron-scalar couplings $\alpha_{e\varphi}=|{\cal Z}^e|^2/4\pi$
comes from stellar cooling through the bremsstrahlung process $e+\ce{^{4}He}\to e+\ce{^{4}He}+\varphi$. It applies to scalar particles sufficiently light to be produced
in stars, typically $M_\varphi$ smaller than  $(1\div 10^3)$ KeV and reads \cite{Raffelt:1996wa}:
\be
\alpha_{e\varphi}<1.4\times 10^{-29}~~~,
\ee
which translates into
\be
|{\cal Z}^e|<1.3\times 10^{-14}~~~.
\ee

In the scalar mass range we are interested in, the strongest constraint on the electron coupling comes from
the existing bounds on the fifth force.
Indeed, if $M_\alpha$ is very small, a long range force between electrons arises, described by the modification to the Newton potential:
\be
\delta V(r)=-\frac{1}{r}\sum_\alpha \frac{\left[{\tt Re}\left({\cal Z}^e_\alpha\right)\right]^2}{4\pi} N_1 N_2 Z_1 Z_2 e^{\dd - M_\alpha r}~~~,
\ee
for two test bodies containing $N_{1,2}$ atoms of atomic numbers $Z_{1,2}$. Here only spin independent interactions induced by the scalar couplings are shown. For pseudoscalars interaction, spin dependent interactions would be induced by the exchange of flavon $\varphi_{\alpha}$ in the nonrelativistic limit. As a consequence, even if the mass of the new particles is very small or exactly zero, they do not mediate a long-range force between unpolarized bodies. Experimental bounds are derived either from tests of the inverse-square law (ISL) or of the
equivalence principle (EP). In the former case the charges of two test bodies are the masses $m_{1,2}=N_{1,2} A_{1,2} u$,
$A_{i}$ being the mass numbers and $u=0.9315$ GeV the atomic mass unit.
Tests of the EP assume charges other than the masses of the two test bodies. Several choices are possible and in our case the charges are given by $N_{1,2} Z_{1,2}$.
The deviations from the Newton potential are parametrized by:
\be
\delta V_{ISL}(r)=-\frac{G m_1 m_2}{r}\alpha e^{\dd - r/\lambda}~~~,
\ee
in tests of ISL, while in the EP case we have:
\be
\delta V_{EP}(r)=-\frac{G m_1 m_2}{r}\tilde\alpha\frac{Z_1}{A_1}\frac{Z_2}{A_2} e^{\dd - r/\lambda}~~~.
\ee
where $G$ is the gravitational constant.
\begin{table}[h!]
\footnotesize
\centering
\resizebox{165mm}{16mm}{
\begin{tabular}{|ccccccccc|}
\hline\hline
\rowcolor{Gray}
$\lambda$ (m)&$10^{-4}$&$10^{-2}$&$10^{2}$&$10^{4}$&$10^{6}$&$10^{8}$&$10^{10}$&$10^{12}$\\
\hline
$M_\varphi/2$ (eV)&$9.87\times10^{-4}$&$9.87\times10^{-6}$&$9.87\times10^{-10}$&$9.87\times10^{-12}$&$9.87\times10^{-14}$&$9.87\times10^{-16}$&$9.87\times10^{-18}$ &$9.87\times10^{-20}$\\
\hline
$\alpha$&$1.07\times10^{-1}$ & $5.02\times 10^{-4}$ & $2.29\times 10^{-3}$ &  $4.76\times 10^{-4}$  & $5.22\times 10^{-6}$ & $6.62\times 10^{-9}$ & $1.06\times 10^{-8}$ & $4.87\times 10^{-8}$\\
\hline
$\tilde\alpha$ &  $-$  &  $-$  &  $1.42\times 10^{-8}$  &   $4.39\times 10^{-9}$ &  $6.11\times 10^{-10}$  &  $7.87\times 10^{-13}$  &  $7.87\times 10^{-13}$  &  $7.87\times 10^{-13}$ \\
\hline
\rowcolor{blue!25}
$|{\tt Re}{\cal Z}^e|$ &$1.98\times 10^{-19}$ & $1.36\times 10^{-20}$ & $3.22\times 10^{-23}$ & $1.79\times 10^{-23}$ & $6.69\times 10^{-24}$ & $2.40\times 10^{-25}$ & $2.40\times 10^{-25}$ & $2.40\times 10^{-25}$\\
\hline\hline
\end{tabular}
}
\caption{Limits on the coupling strength between matter and a scalar particle $\varphi$, $M_\varphi=(\hbar c)/\lambda$,
adapted from refs. \cite{Adelberger:2003zx,Adelberger:2009zz}.
For $M_\varphi<1.55\times 10^{-14}$ eV, the limits on $\tilde\alpha$ and $|{\tt Re}{\cal Z}^e|$
are dominated by the results of the MICROSCOPE experiment \cite{Touboul:2017grn}.
The scalar interaction is assumed to affect electrons only.  To extract the bounds on ${\cal Z}^e$, we have chosen
the representative value $Z_1Z_2/A_1A_2=0.2$.}
\label{tab:limitsISL_update}
\end{table}

\noindent
To make connection with our framework, we assume dominance of the lightest scalar and denote its mass and coupling $M_\varphi=\hbar/(\lambda c)$  and ${\cal Z}^e$, respectively. We can make use of the experimental bounds with the dictionary
\be
\alpha=\frac{Z_1Z_2}{A_1A_2}\frac{[{\tt Re}{\cal Z}^e]^2}{4\pi G u^2}~~~,~~~~~~~~~~~~~~~~~~~~
\tilde\alpha=\frac{[{\tt Re}{\cal Z}^e]^2}{4\pi G u^2}~~~.
\ee
Typical values of the parameter $Z_1Z_2/A_1A_2$ are between 0.16 and 0.22. In table 1 we show the present bounds  \cite{Adelberger:2003zx} on $\alpha$ and $\tilde\alpha$ for some values of $\lambda$ and the corresponding limit on ${\cal Z}^e$, taken as the most restrictive one.

For a fifth force with a range larger than approximately the terrestrial diameter  $\lambda_0\approx 1.27\times 10^7$ m
(corresponding to a scalar mass smaller than about $1.55\times 10^{-14}$ eV, the strongest bound on $\tilde\alpha$
has been set by the MICROSCOPE collaboration \cite{Touboul:2017grn}, that has constrained the E\"otv\"os parameter $\delta(Ti,Pt)=2(a_{Ti}-a_{Pt})/(a_{Ti}+a_{Pt})$ in the range $(-1\pm 13)\times 10^{-15}$, where $a_{Ti,Pt}$ are the free-fall accelerations of the two test bodies in the Earth gravitational field. In our set up and in the limit $\lambda\gg \lambda_0$ the E\"otv\"os parameter is well approximated by:
\be
\delta(Ti,Pt)=\tilde\alpha \left(\frac{Z_{Ti}}{A_{Ti}}-\frac{Z_{Pt}}{A_{Pt}}\right)\frac{Z_{Earth}}{A_{Earth}}~~~.
\ee
From $Z_{Earth}/A_{Earth}=0.4870$ and $(Z_{Ti}/A_{Ti}-Z_{Pt}/A_{Pt})=0.05704$ we get the following 90\% CL limit on
$|\tilde\alpha|$:  $|\tilde\alpha|<7.87\times 10^{-13}$, which translates into
\be
|{\tt Re}{\cal Z}^e|<2.4\times 10^{-25}~~~,
\ee
the strongest
bound to date, for scalar masses below $1.55\times 10^{-14}$ eV~\cite{Fayet:2017pdp,Fayet:2018cjy}.

\subsection{Limits on neutrino coupling}
For very light scalar mediators, limits on scalar-mediated neutrino interactions come mostly from cosmology.
One of the main predictions of the standard cosmological model is the existence of a cosmic background of thermal relic neutrinos.
Weak interactions kept the neutrino background in equilibrium with the cosmological plasma in the early universe. When the temperature of the universe dropped below 1 MeV, neutrinos decoupled and entered the so-called free-streaming regime. This picture is strongly supported by observations.
Such free-streaming regime can be modified if sufficiently strong scalar-mediated neutrino interactions are present.
The modifications depend on the scalar mass $M_\varphi$. If $M_\varphi$ is much larger than the plasma temperature $T$,
scalar exchange can be efficiently modeled by an effective four-neutrino interaction. The characteristic interaction rate $\Gamma$ of scalar-induced neutrino interactions is proportional to $|{\cal Z}^\nu|^4 T^5/M_\varphi^4$, faster than the expansion rate of the universe at high temperatures. This causes a delay of neutrino decoupling and free-streaming, which becomes incompatible with CMB data, unless $|{\cal Z}^\nu|$ is sufficiently small
\cite{Cyr-Racine:2013jua,Archidiacono:2013dua,Lancaster:2017ksf,Oldengott:2017fhy,Kreisch:2019yzn}.
Ref. \cite{Lancaster:2017ksf} obtains the bound:
\be
\frac{|{\cal Z}^\nu|^2}{M_\varphi^2}<(63 ~{\rm MeV})^{-2}\approx   2.16\times 10^7~G_F~~~,
\ee
where ${\cal Z}^\nu$ is assumed to be flavour-independent, and $G_F$ is the Fermi coupling constant.

If $M_\varphi$ is much smaller than the plasma temperature $T$, the rate $\Gamma$ is proportional to $|{\cal Z}^\nu|^4 T$,
smaller than the expansion rate of the universe at high temperatures. Neutrino decouples as in the standard picture, but
when the temperature becomes sufficiently small, neutrino recouple to the cosmological plasma once more and lose their free-streaming.
If this happens too early in the history of the universe, CMB observations are affected. This leads to the bound \cite{Hannestad:2005ex,Archidiacono:2013dua,Forastieri:2015paa,Forastieri:2019cuf}:
\be
|{\cal Z}^\nu_{ii}|<1.2\times 10^{-7}~~~.
\label{CMB}
\ee
Finally, if $\varphi$ is light and decays of the type $\nu_i\to\nu_j \varphi$ are allowed, cosmological observations lead to the limit \cite{Hannestad:2005ex,Archidiacono:2013dua}:
\be
\label{eq:Znu_off}|{\cal Z}^\nu_{ij}|<2.3\times 10^{-11}
\left(\frac{0.05~{\rm eV}}{m}\right)^2~~\,,
\ee
where $m$ is the heavier mass of a given neutrino pair connected by ${\cal Z}^{\nu}_{ij}$.

A scalar particle $\varphi$ interacting with neutrinos can be kept in equilibrium with the universe plasma
during big bang nucleosynthesis (BBN). If full equilibrium is
reached, $\varphi$ would contribute to $N_{eff}$ with $\Delta N_{eff}=4/7$. Even though limits on
$\Delta N_{eff}$ brom BBN are milder than those obtained from CMB observations and still allow for $\Delta N_{eff}=4/7$,
if we require that $\varphi$ does not go in thermal equilibrium before the neutrino decoupling, we get the limit \cite{Huang:2017egl}:
\be
|{\cal Z}^\nu_{ii}|<4.6\times 10^{-6}~~~,
\ee
which is less stringent than the one in eq. (\ref{CMB}).

In summary, strongly interacting neutrinos are compatible with cosmology provided they decouple early enough or recouple late enough.
In our application the lightest scalar $\varphi$ will have a mass smaller than the recombination temperature, about $0.1$ eV
and the relevant bound is the one derived while $M_\varphi<T$, eqs. (\ref{CMB},\ref{eq:Znu_off}). In this we differ from ref. \cite{Smirnov:2019cae}, that applies the stronger
bound $|{\cal Z}^\nu|^2/M_\varphi^2<(3 ~{\rm MeV})^{-2}$, independently of the scalar mass range. This will result
in different conclusions.
\subsection{Limits on ultralight boson masses}
\label{S23}
Limits on ultralight boson masses can also be inferred from purely gravitational systems. Given the smallness of
the gravitational coupling, observable effects can only be expected if some coherent enhancement takes place.
For very light bosons, such an enhancement can occur around a spinning black hole. Bosons can form bound states with the black hole, with an exponentially growing occupation number.
If confined in the vicinity of a Kerr black hole, the boson wave function can extract energy and angular momentum from it, eventually spinning down the black hole. Such superradiance effect is only relevant when the boson Compton wavelength
$\hbar/(M_\varphi c)$ is comparable with the black hole size $\approx G M_{BH}/c^2$:
\be
\frac{G M_\varphi M_{BH}}{\hbar c}\approx 1~~~.
\ee
Considering stellar black holes, $M_{BH}\approx (5\div 50) M_\odot$, and supermassive black holes $M_{BH}\approx (1\div 300)\times 10^6 M_\odot$, the range of boson masses that can be probed is approximately
$(3\div30)\times 10^{-12}$ eV and $(10^{-16}\div 4\times 10^{-19})$ eV, respectively.
By studying rapidly spinning astrophysical black holes, we can potentially exclude or confirm the existence of light massive bosons. The main experimental signatures are the lack of rapidly spinning black holes and monochromatic gravitational
waves that the boson-black hole system can emit either during a transition between two levels or through annihiliation
of bosons into gravitons. Through the observation of spin in stellar black holes, ref. \cite{Arvanitaki:2014wva} have excluded scalar particles with mass in the range:
\be
6\times 10^{-13}~{\rm eV}<M_\varphi<2\times 10^{-11}~{\rm eV}~~~.
\ee
Ref. \cite{Stott:2018opm} analyzed the spin of both stellar and supermassive black holes and have excluded the scalar mass ranges:
\be
7\times 10^{-20}~{\rm eV}<M_\varphi<10^{-16}~{\rm eV}~~~,~~~~~7\times 10^{-14}~{\rm eV}<M_\varphi<2\times 10^{-11}~{\rm eV}~~~.
\label{SR}
\ee
These bounds apply to spin zero particles, independently on their non-gravitational couplings, which are assumed
to be vanishing or negligible. In our study we will adopt the exclusion region in eq. (\ref{SR}).
\section{Scalar NSI from flavour symmetries}
\label{S3}
In this Section we show that scalar NSI naturally arise in models based on flavour symmetries, which aim at an understanding
of fermion masses. In particular, in a large class of such models, it is always possible to cast the relevant part of the Lagrangian
in the form given in eq. (\ref{eq:NSI}).
Our starting point is the lepton sector of a generic flavour model, where all masses are field dependent quantities. Being interested in processes with typical energies
well below the electroweak scale, we set the Higgs multiplet to its vacuum expectation value (VEV) $v/\sqrt{2}$. Instead we keep the full dependence on the
flavon fields $\varphi$, assuming that they are much lighter than the energies relevant to neutrino oscillations.
Throughout this paper we assume Majorana neutrinos, analogous results hold in the case of
Dirac neutrinos. Majorana neutrino masses can arise through the see-saw mechanism or directly from a local
higher dimensional Weinberg operator. Using the two component notation for spinors we have:
\bea
{\cal L}&=&\frac{i}{2}\sum_f\left[\of~ \Kf \osmu \dmu f-(\dmu \of)~ \osmu K^{f\dagger}(\varphi) f \right]+\frac{1}{2}H_{\alpha\beta}(\varphi)\partial_\mu \varphi_\alpha\partial^\mu\varphi_\beta-V(\varphi)\nn\\
&-&\frac{v}{\sqrt{2}} \left[e^c~ {\cal Y}(\varphi) e+\overline e~ {\cal Y}^\dagger(\varphi)\overline{e^c}\right]\label{defining}-\frac{v^2}{2\Lambda_L}\left[\nu~{\cal C}(\varphi) \nu+\overline{\nu}~ {\cal C}^\dagger(\varphi) \overline{\nu}\right]+...~~~,
\eea
where dots denote additional terms related to gauge interactions, to be accounted for in a general
discussion of neutrino oscillations in matter. The matrices $K^{f}(\varphi)+K^{f\dagger}(\varphi)$ $(f=e,e^c,\nu)$ and $H(\varphi)$ are positive definite and $H(\varphi)$ is real symmetric. Flavour indices are understood, ${\cal Y}(\varphi)$ and ${\cal C}(\varphi)$ are complex and ${\cal C}(\varphi)$ is symmetric. They all depend on a set of dimensionless real scalar fields $\varphi_\alpha$. Canonical dimensions can be recovered by redefining $\varphi_\alpha\to \varphi_\alpha/\Lambda$, $\Lambda$ being the characteristic scale of flavour dynamics.

The defining matrices $K^{f}(\varphi)$, $H(\varphi)$, ${\cal Y}(\varphi)$, ${\cal C}(\varphi)$ and the scalar potential
$V(\varphi)$
are constrained by the flavour symmetry of the theory. The latter can be global or local and can be linearly or non-linearly realized. For example,
if the transformations of the flavour symmetry group $G_f$ are global and linearly realized, their action on the fields $f$ and $\varphi$ can be described by:
\be
f\to \Omega_f f~~~,~~~\varphi\to \Omega_\varphi \varphi~~~,
\ee
with unitary ($\Omega_f$) and orthogonal ($\Omega_\varphi$) matrices.
To guarantee invariance under $G_f$, the matrices
$K^{f}(\varphi)$, $H(\varphi)$, ${\cal Y}(\varphi)$ and ${\cal C}(\varphi)$ should satisfy:
\be
\Omega_f^\dagger K^f(\Omega_\varphi\varphi)~ \Omega_f=K^f(\varphi)~~~,~~~~~
\Omega_\varphi^T H(\Omega_\varphi\varphi)~ \Omega_\varphi=H(\varphi)~~~,
\ee
\be
\Omega_{e^c}^T {\cal Y}(\Omega_\varphi\varphi)~ \Omega_e={\cal Y}(\varphi)~~~,~~~~~
\Omega_\nu^T {\cal C}(\Omega_\varphi\varphi)~ \Omega_\nu={\cal C}(\varphi)~~~.
\ee
The scalar potential $V(\varphi)$ obeys:
\be
V(\Omega_\varphi\varphi)=V(\varphi)~~~.
\ee
If $\Kf=H(\varphi)=\mathbb{1}$, the kinetic terms are canonical. This is not the most general case and in general flavour symmetries allow for non canonical kinetic terms. If the flavour symmetry is continuous and local, there are
additional gauge interactions beyond the SM ones. The associated gauge bosons are expected to mediate flavour changing neutral currents and here we assume they are sufficiently heavy and do not play any role in neutrino oscillations
in matter. If the symmetry is non-linearly realized, as for the case of the modular group, the matrices $K^{f}(\varphi)$, $H(\varphi)$, ${\cal Y}(\varphi)$, ${\cal C}(\varphi)$ and the scalar potential $V(\varphi)$ have to satisfy properties which will be specified in concrete examples.

We are interested in Yukawa trilinear interactions of the scalar particles with neutrinos and with electrons. To analyze them, we proceed through a series of standard steps. In detail,
we expand the functions $\Kf$ $(f=e,e^c,\nu)$, $H(\varphi)$, $V(\varphi)$, ${\cal Y}(\varphi)$ and ${\cal C}(\varphi)$ around the VEVs $\varphi^0_\alpha$, to the first order. Then we move to a basis where the fermion fields have canonical kinetic terms. We use the equation of motion to cast all interactions in the Yukawa form. Finally, we move to the mass eigenstate basis for the scalar fields and for the charged leptons.
We get:
\bea
{\cal L}&=&i\sum_{f=e,e^c,\nu}\of~\osmu \dmu f+\frac{1}{2}\partial_\mu \varphi_\alpha\partial^\mu\varphi_\alpha-\frac{1}{2}\varphi_\alpha M_{\alpha\beta}^2 \varphi_\beta\label{bel}\\
&-& e^c(m_e+{\cal Z}^e_\alpha \varphi_\alpha)e -\frac{1}{2}\nu(m_\nu+{\cal Z}^\nu_\alpha\varphi_\alpha)\nu+h.c.+...\,, \nn
\eea
where the matrices $m_e$ and $M^2_{\alpha\beta}=M^2_\alpha \delta_{\alpha\beta}$ are diagonal and positive definite:
\bea
m_e&=&U^{T}_{e^c}m_e' U_e~~,~~~~~ m_\nu=U_e^T m_\nu' U_e~~,~~~~~M^2=\Omega^T {M'}^2 \Omega~~~,
\nn\\
&&\nn\\
{\cal Z}^e_\alpha&=&U_{e^c}^T {{\cal Z}^e_\gamma}' U_e~\Omega_{\gamma \alpha}~~~,~~~~~~~{\cal Z}^\nu_\alpha=U_{e}^T {{\cal Z}^\nu_\gamma}' U_e~\Omega_{\gamma \alpha}~~~.
\eea
Notice that we made on neutrinos the same transformation $U_e$ as in the left-handed charged lepton sector so that in this basis the neutrino mass matrix is diagonalized by the PMNS matrix.
The primed matrices refer to the basis where all fields are canonically normalized. They are given by:
\bea
\label{surprise}
m_e'&=&\frac{v}{\sqrt{2}}{\cal Y}'_0~~~,~~~~~~~m_\nu'=\frac{v^2}{\Lambda_L}{\cal C}'_0\,,\nn\\
{{\cal Z}^e_\alpha}'&=&\frac{v}{\sqrt{2}}\left[{\cal Y}'_{\alpha 0}-\frac{1}{2}\left({\cal Y}'_0 {K'}^{e\dagger}_{\alpha0}+{{K'}^{e^c}_{\alpha0}}^* {\cal Y}'_0\right)\right]\,,\\
{{\cal Z}^\nu_\alpha}'&=&\frac{v^2}{\Lambda_L}\left[{\cal C}'_{\alpha 0}-\frac{1}{2}\left({\cal C}'_0 {K'}^{\nu\dagger}_{\alpha0}+{{K'}^{\nu}_{\alpha0}}^* {\cal C}'_0\right)\right]\,,\nn\\
{M'}^2_{\alpha\beta}&=&V'_{\alpha\beta0}~~~,\nn
\eea
where ${K'}^{f}_{\alpha0}$, ${\cal Y}'_0$, ${\cal C}'_0$, ${\cal Y}'_{\alpha 0}$, ${\cal C}'_{\alpha 0}$ and $V'_{\alpha\beta0}$
are built in the following way. Starting from the defining Lagrangian, eq. (\ref{defining}), we
expand the functions $K^f(\varphi)$, ${\cal Y}(\varphi)$, ${\cal C}(\varphi)$, $H_{\alpha\beta}(\varphi)$ and $V(\varphi)$
around the minimum $\varphi^0_\alpha$ of the scalar potential $V(\varphi)$:
\bea
\Kf&=&K^f_0+K^f_{\alpha 0}~\varphi'_\alpha+...\,,~~~~
{\cal Y}(\varphi)={\cal Y}_0+{\cal Y}_{\alpha 0}~\varphi'_\alpha+...\,,~~~~
{\cal C}(\varphi)={\cal C}_0+{\cal C}_{\alpha 0}~\varphi'_\alpha+...\,,\nn\\
&&\nn\\
H_{\alpha\beta}(\varphi)&=&H_{\alpha\beta0}+...\,,~~~~~~~~~~~~
V(\varphi)=V_0+V_{\alpha\beta 0}~\varphi'_\alpha \varphi'_\beta+...\,,~~~~~
\eea
where $\varphi'_\alpha=\varphi_\alpha-\varphi^0_\alpha$ and we use the notation $K^f_0=K^f(\varphi^0)$, $K^f_{\alpha 0}=(\partial K^f/\partial\varphi_\alpha)(\varphi^0)$ and similarly for the other quantities. We put kinetic terms in a canonical form through a
combination of a unitary matrix $T$ and a rescaling $(D_0)^{-1/2}$:
\bea
&&\frac{1}{2}(D^f_0)^{-1/2} {T^f}^\dagger (K^{f}_{0}+K^{f \dagger}_{0})~ T^f (D^f_0)^{-1/2}=\mathbb{1}~~~,\nn\\
&&(D^\varphi_0)^{-1/2} {T^\varphi}^T~ H_0~ T^\varphi (D^\varphi_0)^{-1/2}=\mathbb{1}~~~.
\eea
Finally, we define the primed quantities by moving to the basis where kinetic terms are canonically normalized:
\bea
{K'}^{f}_{\alpha0}&=&(D^f_0)^{-1/2} {T^f}^\dagger K^{f}_{\alpha0} T^f (D^f_0)^{-1/2}\,,\nn\\
{\cal Y}'_0&=&(D^{e^c}_0)^{-1/2} {T^{e^c}}^T {\cal Y}_0~ T^e (D^e_0)^{-1/2}\,,\nn\\
{\cal C}'_0&=&(D^{\nu}_0)^{-1/2} {T^{\nu}}^T {\cal C}_0~ T^\nu (D^\nu_0)^{-1/2}\,,\\
{\cal Y}'_{\alpha 0}&=&(D^{e^c}_0)^{-1/2} {T^{e^c}}^T {\cal Y}_{\alpha 0}~ T^e (D^e_0)^{-1/2}\,,\nn\\
{\cal C}'_{\alpha 0}&=&(D^{\nu}_0)^{-1/2} {T^{\nu}}^T {\cal C}_{\alpha 0}~ T^\nu (D^\nu_0)^{-1/2}\,,\nn\\
V'_{\alpha\beta 0}&=&[(D^\varphi_0)^{-1/2} {T^\varphi}^T]_{\alpha\alpha'}~ V_{\alpha'\beta' 0}~ [T^\varphi (D^\varphi_0)^{-1/2}]_{\beta'\beta}~~~.\nn
\eea
The fields undergo the overall transformation (to simplify the notation here $\varphi$ stands for the
fluctuation $\varphi-\varphi^0$):
\be
f\to T^f(D^f_0)^{-1/2}U_f~ f~~~,~~~~~\varphi\to T^\varphi (D^\varphi_0)^{-1/2} \Omega~ \varphi~~~.
\ee
In particular we are interested in the interaction with the electron (first generation charged lepton) and the Lagrangian of eq. (\ref{bel}) specializes as follows:
\bea
{\cal L}&=&i\sum_{f=e,e^c,\nu}\of~\osmu \dmu f+\frac{1}{2}\partial_\mu \varphi_\alpha\partial^\mu\varphi_\alpha-\frac{1}{2}\varphi_\alpha M_{\alpha\beta}^2 \varphi_\beta\label{bel1}\\
&-& e^c_1\left[(m_e)_{11}+\left({\cal Z}^e_\alpha\right)_{11} \varphi_\alpha\right]e_1 -\frac{1}{2}\nu(m_\nu+{\cal Z}^\nu_\alpha\varphi_\alpha)\nu+h.c.+... \nn
\eea
and coincides with that of eq. (\ref{eq:NSI}). From eq. (\ref{surprise}) we see that in this class of models scalar NSI arise not only from the field-dependence of Yukawa couplings, but also from non-canonical kinetic terms allowed by the flavour symmetry. These give rise to additional interaction terms between leptons and scalars, which have to be properly included to analyze the impact of scalar exchange.

\section{Models with abelian flavour symmetries}
\label{S4}
We discuss here two models of lepton masses based on continuous abelian flavour symmetries.
We first analyze a very simple model, to illustrate the difficulties arising when looking for observable
effects generated by scalar NSI. Then we move to a more complex model, where the prospects of a
detectable signal are more promising.
\subsection{A toy model}
It is instructive to analyze a simple model with an abelian flavour symmetry group U(1). Lepton doublets of the three generations are assigned a common charge $q/2$,
while the overall charge of the bilinear $e^c_1 e_1$ is denoted by $p$. Both $q$ and $p$ are positive integers. We neglect intergenerational mixing in the charged lepton sector and we consider canonical kinetic terms, to start with. If the symmetry is spontaneously broken by a single flavon $\varphi$, carrying a negative unite of the abelian charge, the relevant Lagrangian reads:
\be
{\cal L}=\dd i\sum_{f=e^c_1,e_1,\nu} \of~ \osmu \dmu f
-\left[\frac{y_0 v}{\sqrt{2}} \left(\frac{\varphi}{\Lambda}\right)^p e^c_1~ e_1
+\frac{v^2}{2\Lambda_L} \left(\frac{\varphi}{\Lambda}\right)^q \nu~{\cal C}_0~ \nu+h.c.\right]~~~,
\ee
where $y_0$ and the matrix elements ${\cal C}_{0ij}$ are of order one (it is not restrictive to assume $y_0>0$), $\Lambda_L$ is the scale associated to the breaking of $B-L$ and $\Lambda$ is the cutoff scale. When $\varphi$ acquires the VEV $\varphi_0$, mass and interaction terms are generated:
\be
m_{e_1}=\frac{y_0 v}{\sqrt{2}} \left(\frac{\varphi_0}{\Lambda}\right)^p~,~~~~
m_\nu=\frac{v^2}{\Lambda_L} \left(\frac{\varphi_0}{\Lambda}\right)^q{\cal C}_0~,~~~~
{\cal Z}^e=\frac{1}{\sqrt{2}}~p~\frac{m_{e_1}}{\varphi_0}~,~~~~
{\cal Z}^\nu=\frac{1}{\sqrt{2}}~q~\frac{m_\nu}{\varphi_0}~~.
\label{coupl}
\ee
With the above charge assignment, the mass matrix for light neutrinos is of anarchical type, compatible with present data. The field $\varphi$ is complex and both scalar and pseudoscalar interactions are induced\footnote{A factor of $1/\sqrt{2}$ accounts for the real scalar component in the coupling constants ${\cal Z}^{e,\nu}$.}.
If the U(1) symmetry is local, the pseudoscalar component of $\varphi$ gets eaten by the gauge vector boson via the Higgs mechanism. The scalar component of $\varphi$ describes a physical particle which can be very light. For instance, in a supersymmetric realization, $\varphi$ can parametrize a nearly flat direction, with a resulting very light scalar degree of freedom and a large VEV $\varphi_0$. The latter might help suppressing both the coupling ${\cal Z}^e$, as we see from eq. (\ref{coupl}), and the interaction induced by gauge vector boson exchange.
We denote $M_\varphi$ the mass of the scalar particle.
\begin{figure}[h!]
\centering
\includegraphics[width=0.65\textwidth]{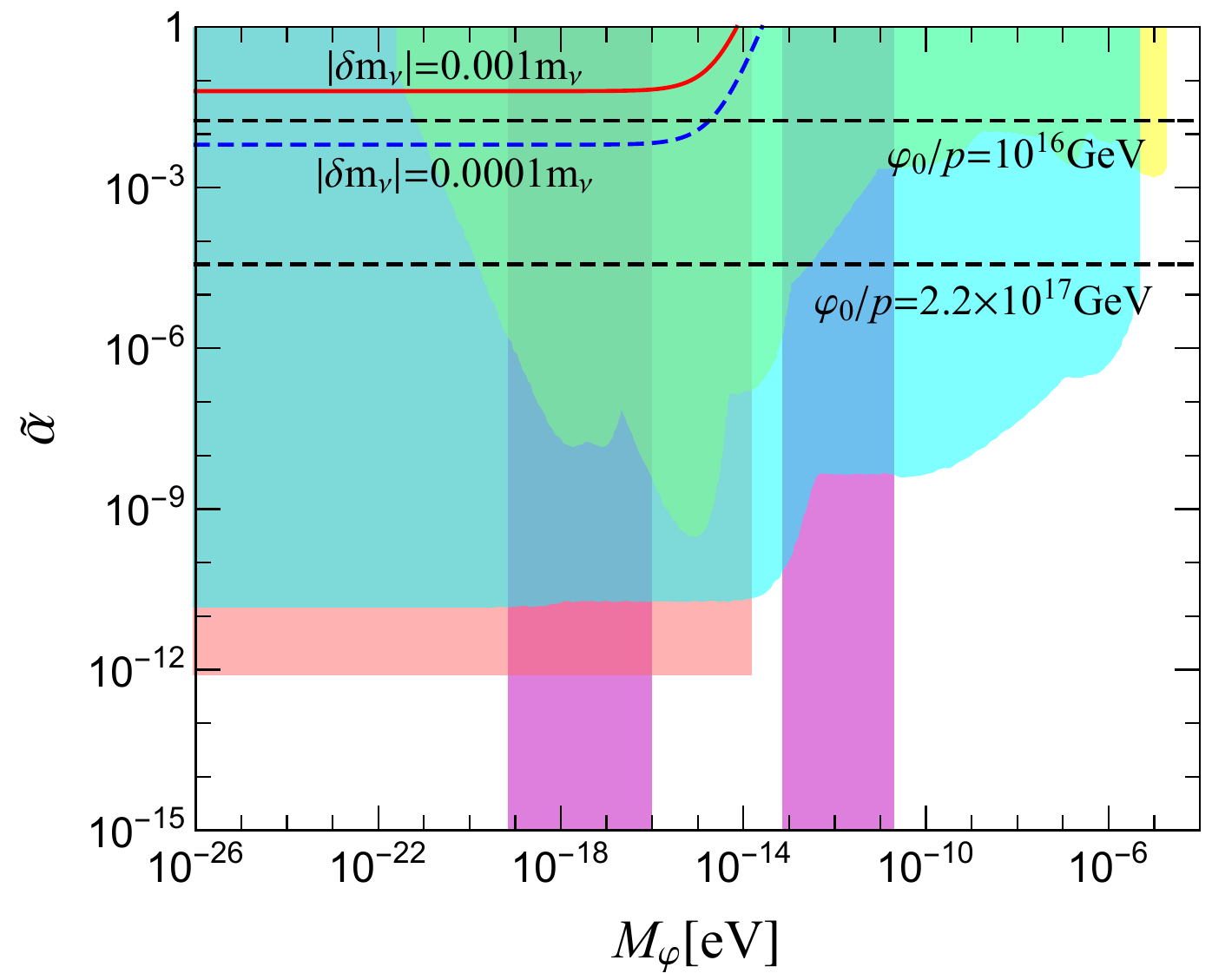}
\caption{The red(blue) contour shows $|\delta m_\nu(0)/m_\nu|=0.001(0.0001)$, for $q/p=1$. The combination
$n_e^0 F(M_\alpha R)$ is the one of eq. (\ref{eq:FF}), with $n_e(r)$ from ref. \cite{Bahcall:2004pz,Bahcall_web}. The vertical bands in purple are excluded from black hole superradiance. The other colored regions are excluded by tests of the Newton law: yellow from test of ISL and blue and pink from tests of the EP for interaction range larger than 1cm, see the discussion in Sections 2.1 and 2.3. Also shown are two dashed contours of $p/\varphi_0$.}
\label{contour0}
\end{figure}
\vskip 0.5 cm

As discussed in Section \ref{LRF},  the interactions of a very light scalar particle are severely constrained.
In the model under discussion the parameters $\tilde\alpha$ and $\lambda$ probed by
the experiments testing long range forces are given by:
\be
\tilde\alpha=\frac{p^2m_{e_1}^2}{8\pi G u^2\varphi_0^2}~~~,~~~~~
\lambda=\frac{\hbar}{M_\varphi c}~~.
\label{talfa}
\ee
The shift $\delta m_\nu(0)$ of the neutrino mass matrix at the center of a spherical region of radius $R$ with spherically symmetric electron density $n_e(r)$ is given by:
\be
\delta m_\nu(0)=-n_e^0 \frac{pq~ m_{e_1} m_\nu}{2\varphi_0^2}\frac{F(M_\varphi R)}{M^2_\varphi}~~~,
\ee
where the combination $n_e^0 F(M_\varphi R)$ is given in eq. (\ref{eq:FF}).
We can replace the dependence on $\varphi_0$ by that on $\tilde\alpha$ by making use of eq. (\ref{talfa}) and obtain:
\be
\label{eq:mnu_tm1}\delta m_\nu(0)=-4\pi n_e^0 G u^2\tilde\alpha \frac{q~ m_\nu}{p~m_{e_1}}\frac{F(M_\varphi R)}{M^2_\varphi}~~~.
\ee
In fig. \ref{contour0}, in the plane $(M_\varphi,\tilde\alpha)$, we show contours corresponding to $|\delta m_\nu(0)/m_\nu|=0.001$ and $|\delta m_\nu(0)/m_\nu|=0.0001$, probably below threshold for observation with the present accuracy. We have chosen $q/p=1$ and $R\approx 6.955\times 10^5$ Km, to estimate the effect
in the sun.
We see that not even extremely small scalar masses $M_\varphi$ allow to satisfy the bound on $\tilde\alpha$ and, at the same time,
to produce a sizable effect in $\delta m_\nu(0)$. This is due to the finite region where matter effects take place, at the origin of the cutoff $F(M_\varphi R)$ in $\delta m_\nu(0)$ and responsible for the flat behavior of the red curve in fig. \ref{contour0}.
We also see that, to deplete $|{\cal Z}^e|$ below the present upper bound, we would need a value of $\varphi_0/p$ much larger than the Planck scale. Essentially no room for an observable effect is left by the existing constraints in this model.

If we turn on non-canonical kinetic terms, the picture remains qualitatively unchanged. The U(1) symmetry allows the kinetic functions
\be
K^{e(e^c)}(\varphi)=1+b^{e(e^c)} \frac{|\varphi|^2}{\Lambda^2}+...~~~,~~~~~(K^\nu)_{ij}(\varphi)=\delta_{ij}+b^\nu_{ij}\frac{|\varphi|^2}{\Lambda^2}+...\,.
\ee
Here the $b^f$ coefficients are generically of order one and dots stand for higher order contributions in the $|\varphi|^2/\Lambda^2$ expansion.
We see that the effect of the new terms is to modify the effective couplings ${\cal Z}^e$ and ${\cal Z}^\nu$ by subleading contributions. We now have:
\bea
{\cal Z}^e&=&\frac{1}{\sqrt{2}}\left[p~\frac{\hat m_{e_1}}{\varphi_0}-(b^e+b^{e^c})\frac{\hat m_{e_1}}{\varphi_0}\left(\frac{\varphi_0^2}{\Lambda^2}\right)+...\right]\nn\\
{\cal Z}^\nu&=&\frac{1}{\sqrt{2}}\left[q~\frac{\hat m_\nu}{\varphi_0} - \left(\frac{\hat m_\nu}{\varphi_0} b^{'\nu}+b^{'{\nu}T} \frac{\hat m_\nu}{\varphi_0}\right)\left(\frac{\varphi_0^2}{\Lambda^2}\right) +...\right]
\eea
Here $b^{'\nu}$ is also a matrix with generic, order one entries. The new contributions are subleading, unless $p$ and/or $q$ vanish. To suppress the electron-scalar interaction we would need $p=0$, but in this case
the electron mass would be adjusted by hand and not explained by the symmetry.

We could also contemplate the possibility of a mixing between $\varphi$ and the Higgs particle $h$. The lepton masses are those of eq. (\ref{coupl}), while the couplings ${\cal Z}^{e,\nu}_\varphi$ are obtained by the replacement:
\be
p\to p\cos\theta-\frac{\sqrt{2}\varphi_0}{v}\sin\theta~~~,~~~~~~~q\to q\cos\theta-\frac{\sqrt{2}\varphi_0}{v}\sin\theta~~~.
\ee
Here $\theta$ denotes the mixing angle between interaction and mass eigenstates in the  $(\varphi,h)$ sector.
For any value of $p$ and $\varphi_0$, we can look for an angle $\theta$ such that ${\cal Z}^e_\varphi$ is
reduced to the tiny value $10^{-25}$. For example, if $\varphi_0\approx10^{10}$ GeV and $p$ is of order one, we need an angle $\theta\approx 10^{-8}$, tuned to an extremely good precision to achieve the desired cancellation.
In particular, while in this example $p/\sqrt{2}\varphi_0$ and $\sin\theta/v$ are both individually of order $10^{-10}$ GeV$^{-1}$,  their difference is required to be twelve order of magnitudes smaller. If such a miraculous cancellation takes place,
by choosing $M_\varphi=10^{-16}$ eV we would obtain $\delta m\approx\sqrt{\Delta m^2_{atm}}$.

We conclude that, within U(1) models with a single flavon, observable effects induced by scalar NSI can only occur at the price of a severe fine tuning.
\subsection{A variant}
In this section we show that in abelian flavour models it is possible to achieve observable effects.
We consider a model invariant under the abelian symmetry U(1)$_1\times$U(1)$_2$. Lepton doublets of the three generations are neutral under U(1)$_1$ and have a common charge $q/2$ under U(1)$_2$,
while the bilinear $e^c_1 e_1$, neutral under U(1)$_2$, have an overall charge $p$ under U(1)$_1$ ($q$ and $p$ are positive integers as before). This can be realized via the charge assignment shown in table \ref{variant}.

\begin{table}[h!]
\footnotesize
\centering
\begin{tabular}{|cccccc|}
\hline\hline
\rowcolor{Gray}
&$e^c$&$l$& {\tt Higgs}& $\varphi_1$& $\varphi_2$\rule[-2ex]{0pt}{-4ex}\\
\hline
U(1)$_1$&$p$&$0$&$0$&$-1$&$0$\rule[-1ex]{0pt}{2ex}\\
\hline
U(1)$_2$&$-q/2$&$q/2$&$0$&$0$&$-1$\rule[-1ex]{0pt}{2ex}\\
\hline \hline
\end{tabular}
\caption{Charge assignement for a model invariant under U(1)$_1\times$U(1)$_2$. }
\label{variant}
\end{table}

\begin{figure}[h!]
\centering
\includegraphics[width=0.7\textwidth]{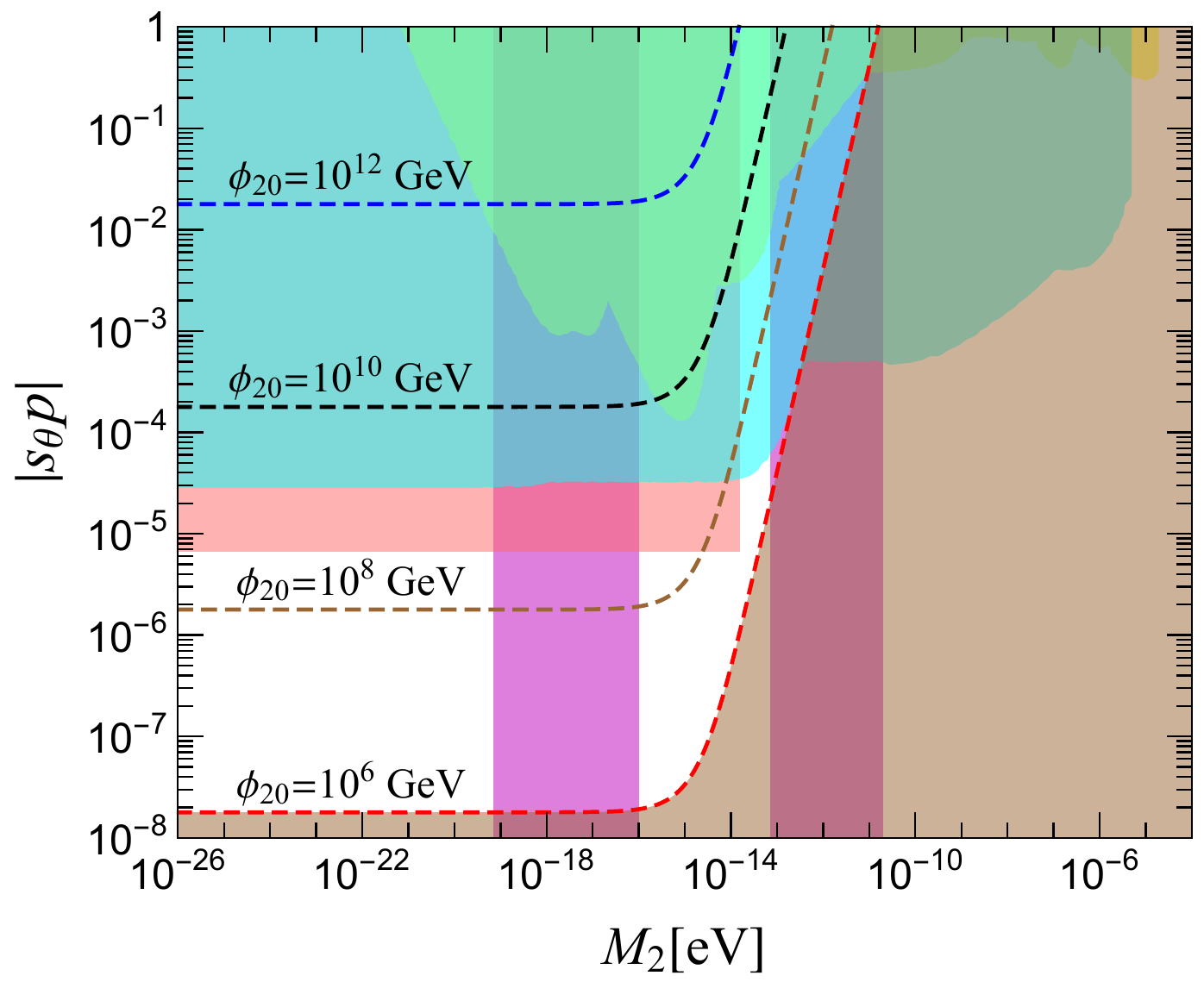}
\caption{For fixed values of $\varphi_{10}=10^{16}$ GeV and $\phi_{20}\equiv 2\varphi_{20}/(q c_{\theta})$, contours $|\delta m_\nu(0)/m_\nu|=0.1$. The combination $n_e^0 F(M_\alpha R)$ is the one of eq. (\ref{eq:FF}), with $n_e(r)$ from ref. \cite{Bahcall:2004pz,Bahcall_web}.
The vertical bands in purple are excluded from black hole superradiance. The other colored regions are excluded by tests of the Newton law: yellow from test of ISL and blue and pink from tests of the EP, see the discussion in Sections 2.1 and 2.3. In the brown region $\Lambda_L<1$ TeV, see text.}
\label{contour1}
\end{figure}

We assume here canonical kinetic terms. Even though the flavour symmetry allows for non-canonical contributions, these would not play a dominant role in a large portion of the parameter space. At the same time, by allowing for extra parameters, they would obscure our discussion. In this limit we have:
\be
{\cal Y}(\varphi)=y_0\left(\frac{\varphi_1}{\Lambda}\right)^p~,~~~~{\cal C}(\varphi) =\left(\frac{\varphi_2}{\Lambda}\right)^q{\cal C}_0~,
\ee
giving rise to masses:
\be
m_{e_1}=\frac{y_0 v}{\sqrt{2}} \left(\frac{\varphi_{10}}{\Lambda}\right)^p~~~,~~~~~
m_\nu=\frac{v^2}{\Lambda_L} \left(\frac{\varphi_{20}}{\Lambda}\right)^q{\cal C}_0~~~.
\ee
Also in this case the pseudoscalar components of $\varphi_{1,2}$ are eaten up by the gauge vector bosons of U(1)$_1\times$U(1)$_2$,
assumed to be very heavy. Denoting by $\theta$ the mixing angle between mass and interaction bases
in the scalar sector, we have (here $s_\theta\equiv\sin\theta$ and $c_\theta\equiv\cos\theta$):
\be
{\cal Z}^e_1=p~\frac{c_\theta}{\sqrt{2}}\frac{m_{e_1}}{\varphi_{10}}~,~~
{\cal Z}^e_2=p~\frac{s_\theta}{\sqrt{2}}\frac{m_{e_1}}{\varphi_{10}}~,~~
{\cal Z}^\nu_1=-q~\frac{s_\theta}{\sqrt{2}}\frac{m_\nu}{\varphi_{20}}~,~~
{\cal Z}^\nu_2=q~\frac{c_\theta}{\sqrt{2}}\frac{m_\nu}{\varphi_{20}}~~.
\ee
The modification of the Newton potential due to scalar exchange is:
\be
\label{eq:pot_toy2}\delta V(r)=-\frac{N_1 N_2 Z_1 Z_2}{4\pi r}\left[({\cal Z}^e_1)^2 e^{\dd - M_1 r}+({\cal Z}^e_2)^2 e^{\dd - M_2 r}\right]~~~,
\ee
where $M_{1,2}$ are the scalar masses.
To evade the bounds coming from long range forces, while leaving room for sizable scalar NSI we assume $M_1\gg M_2$. We look for a region of the parameter space where $\varphi_1$ is sufficiently heavy not to appreciably contribute to $\delta V(r)$, and $\varphi_2$ is sufficiently light to induce significant scalar NSI effects.
For instance, for $M_1\ge10^{-4}$ eV and $\varphi_{10}=10^{16}$ GeV, the contribution of the scalar $\varphi_1$
to $\delta V(r)$ is beyond the accuracy of the present test of ISL and EP.  In this region of parameter space we have:
\be
\tilde\alpha=\frac{p^2 s_\theta^2m_{e_1}^2}{8\pi G u^2\varphi_{10}^2}~~~,~~~~~
\lambda=\frac{\hbar}{M_2 c}~~~.
\label{talfav}
\ee
The shift $\delta m_\nu(0)$ of the neutrino mass matrix at the center of a spherical region of radius $R$ with spherically symmetric electron number density $n_e(r)$ is given by:
\be
\delta m_\nu(0)=-n_e^0  \frac{s_\theta c_\theta~pq~ m_{e_1} m_\nu}{2\varphi_{10}\varphi_{20}}\left[\frac{F(M_2 R)}{M_2^2}-\frac{F(M_1 R)}{M_1^2}\right]~~~,
\ee
where the combinations $n_e^0 F(M_{1,2} R)$ are given in eq. (\ref{eq:FF}).
To estimate the observability of such an effect, we work in the region $M_1\ge10^{-4}$ eV and $\varphi_{10}=10^{16}$ GeV, where the contribution to $\delta m_\nu(0)$ from $\varphi_1$ exchange is negligible.
Then the neutrino mass shift can be expressed as
\begin{equation}
\label{eq:mnu_tm2}\delta m_\nu(0)=\pm\frac{n_e^0 F(M_2 R)}{\phi_{20}M^2_2}  \sqrt{8\pi G u^2\tilde{\alpha}}~ m_{\nu}\,,
\end{equation}
with $\phi_{20}\equiv 2\varphi_{20}/(qc_{\theta})$. As in the previous case, we analyze the effect induced by the sun, taking $R\approx 6.955\times 10^5$ Km. Since $\varphi_{10}$ is fixed, from eq. (\ref{talfav}) the bounds on $\tilde\alpha$ can be directly translated in bounds on the combination $s_\theta p$, shown in fig. \ref{contour1}. In the plane $(M_2,s_\theta p)$ we display contours corresponding to $|\delta m_\nu(0)/m_\nu|=0.1$, which we tentatively take as threhsold for observability, for several choices of the combination $\phi_{20}$.

\vskip 0.5 cm
We see that for $M_2<7\times10^{-14}$ eV, for sufficiently small $s_\theta$
and $\varphi_{20}$, there can be room for detectable effects in neutrino oscillations due to
scalar NSI mediated by flavons.
We exploited the fact that the bound coming from long range forces depend on $\varphi_{10}$ but not on $\varphi_{20}$. We can maximize $\varphi_{10}$ while lowering $\varphi_{20}$ to enhance $\delta m_\nu$.
However, the VEV $\varphi_{20}$ cannot be arbitrarily small. Indeed $m_\nu\approx (\varphi_{20}/\Lambda)^q v^2/\Lambda_L$, and the requirement of having the scale of breaking of the lepton number $\Lambda_L$ larger than 1 TeV leads to $\varphi_{20}>10^{6}$ GeV for $q=1$ and $\varphi_{20}>10^{12}$ GeV for $q=2$, when assuming $\Lambda=10^{18}$ GeV. In fig. \ref{contour1}, this bound is represented by $\phi_{20}>10^{6}$ GeV since $\phi_{20}\simeq2\varphi_{20}/q$ for small $s_{\theta}$ and it is of the same order of magnitude as $\varphi_{20}$.

The region of parameter space allowing detectable effects via scalar NSI needs some amount of fine tuning. Indeed, while $M_{1,2}$ and $s_\theta$ are free input parameters,
avoiding fine tuning to achieve $M_2\ll M_1$ requires the approximate relation $M_1\approx M_2/s_\theta$. This can be derived by the most general real symmetric 2$\times$2 mass matrix in the scalar sector:
\be
\left(
\begin{array}{cc}
m^2  &  \mu^2\\
\mu^2 & M^2
\end{array}
\right)~~~,
\ee
with $(m^2,\mu^2)\ll M^2$. We have $M_1^2\approx M^2$, $M_2^2\approx m^2-\mu^4/M^2$ and $s_\theta\approx \mu^2/M^2$. Fine tuning is avoided if the smallest eigenvalue $M_2^2$ does not require a precise cancellation between $m^2$ and $\mu^4/M^2$, that is $M_2^2\approx \mu^4/M^2$ or $M_1\approx M_2/s_\theta$. In our model it is not possible to satisfy this relation. Indeed, from fig. \ref{contour1} we see that the
ratio $M_2/s_\theta p$ is typically much smaller than $M_1\ge 10^{-4}$ eV, assumed to escape limits from
long range forces due to $\varphi_1$ exchange. We conclude that a considerable cancellation should take place between $m^2$ and $\mu^4/M^2$, to reproduce a small $M_2$.

Barring naturalness considerations, this model shows that observable effects in matter neutrino oscillations associated to scalar NSI as predicted by flavour models are indeed possible. They require a very light scalar degree of freedom with a tiny coupling to electrons, due to the extremely strong bounds on long range forces. A comparatively larger coupling to neutrinos is needed to achieve observability. In the above model these ingredients are related to the different VEVs of the scalars responsible for electron and neutrino masses and to the small mixing between the two.

\section{Modular invariant models}
\label{S5}

In this section we shortly review the formalism of supersymmetric modular invariant theories \cite{Ferrara:1989bc,Ferrara:1989qb} applied to flavour physics \cite{Feruglio:2017spp} and we derive the linearized fermion-modulus interactions.
The Lagrangian $\mathscr{L}$ depends on a set of chiral supermultiplets $\phi$ comprising the modulus $\phi_1 =\Lambda\tau$ (${\tt Im} \tau>0$) and other superfields $\phi_i$ ($i>1)$:
\be
\mathscr{L}=\int d^2\theta d^2\bar\theta~ K(\phi,\bar{\phi})+\int d^2\theta~ W(\phi)+\int d^2\bar\theta~ {\overline W}(\bar\phi)~~~.
\ee
The Lagrangian is invariant under transformations $\gamma$ of the homogeneous modular group $\Gamma=SL(2,Z)$:
\be
\tau\to\gamma\tau=\frac{a \tau+b}{c\tau+d}~~~,~~~~~\phi_i\to (c\tau+d)^{-k_i}\rho(\gamma)_{ij} \phi_j~~~~~(i,j>1)~~~.
\ee
where $a$, $b$, $c$, $d$ are integers obeying $ad-bc=1$ and $\rho(\gamma)$ is a unitary representation of
the group $\Gamma_N'=\Gamma/\Gamma(N)$, obtained as a quotient between the group $\Gamma$ and a principal congruence subgroup $\Gamma(N)$, the integer $N$ being the level of the representation\footnote{Following ref. \cite{Liu:2019khw}, we consider here homogeneous finite modular groups $\Gamma_N'$ instead of their inhomogeneous counterpart $\Gamma_N$.}.
In general $\rho(\gamma)$ is a reducible representation and all superfields
belonging to the same irreducible component should have the same weight $k_i$. Some of the superfields $\phi_i$ $(i>1)$
may describe flavons, gauge singlets with the scalar component acquiring a large VEV $\langle\phi_i\rangle$. We adopt a minimal K$\ddot{\mathrm{a}}$hler potential:
\begin{equation}
\label{eq:Kal}K(\phi,\bar{\phi}) = -h \Lambda^2\log(-i\tau+i\overline{\tau})+ \sum_{i>1}(-i\tau+i\overline{\tau})^{-k_i}\overline{\phi}_{\bar{i}}\phi_i\,.
\end{equation}
In the following, we denote by $(\phi_i, \psi_i)$ the spin-(0, 1/2) components of the chiral superfields $\phi_i$ \footnote{The distinction between superfields and their scalar components should be clear from the context.}.
The terms bilinear in the fermion fields read~\cite{Brignole:1996fn}:
\begin{equation}
\mathscr{L}_{F}=\mathscr{L}_{F, K}+\mathscr{L}_{F, 2}\,,
\end{equation}
with
\begin{equation}
\label{eq:Lag_modu}\mathscr{L}_{F, K}=iK_{\bar{j}i}\overline{\psi}^{\bar{j}}\bar{\sigma}^{\mu}D_{\mu}\psi^{i}\,,~~~~~
\mathscr{L}_{F, 2}=
-\frac{1}{2}\left[W_{ij} - W_l (K^{-1})^{l \bar{m}}
K_{\bar{m} i j }\right]\psi^{i}\psi^{j}+h.c.\,,
\end{equation}
where unbarred(barred) indices in $K$ and $W$ stand for derivatives with respect to holomorphic (anti-holomorphic) fields.
The covariant derivative is:
\be
D_{\mu}\psi^{i}=\partial_{\mu}\psi^i+\left(K^{-1}\right)^{i\bar{m}}K_{\bar{m}kl}\partial_{\mu}\phi^k
\psi^l~~~.
\ee
With our choice of the K$\ddot{\mathrm{a}}$hler potential, eq. (\ref{eq:Kal}), the K$\ddot{\mathrm{a}}$hler matrix $K_{\bar{i}j}$
reads:
\be
K_{\bar{i}j}=
\left(
\begin{array}{cc}
\frac{h}{\xi^2}+\dd\sum_{l>1}\frac{k_l(k_l+1)}{\xi^{k_l+2}\Lambda^2}\,\overline{\phi}_l\phi_l&-\frac{ik_j}{\xi^{k_j+1}\Lambda}\overline{\phi}_{\bar{j}}\\
\frac{ik_i}{\xi^{k_i+1}\Lambda}\phi_i&\frac{1}{\xi^{k_i}}\delta_{ij}
\end{array}
\right)~~~,
\ee
where we define $\xi=-i\tau+i\overline{\tau}$. We assign vanishing weight to the fields that
acquire a non vanishing VEV, such as the flavons. Under this assumption, when fields are set to their VEVs, the K$\ddot{\mathrm{a}}$hler matrix $K_{\bar{i}j}$ is diagonal. Such a case can be easily generalized, without affecting
most of our considerations. The transformation that makes the kinetic terms canonical is:
\be
\phi_1\to\frac{\langle\xi\rangle}{\sqrt{h}}\phi_1~~~,~~~~~\phi_i\to\langle\xi\rangle^{k_i/2}\phi_i,~~~~~\psi_i\to\langle\xi\rangle^{k_i/2}\psi_i~~~i>1,
\ee
where $\langle\xi\rangle$ stands for the VEV of $\xi$.

By expanding the Lagrangian around the VEV $(\langle\tau\rangle, \langle\phi_i\rangle)$, after rescaling the fields to make the kinetic term canonical, we get:
\begin{eqnarray}
\label{eq:Lag_sim1}
\mathscr{L}&=&i\overline{\psi}\bar{\sigma}^{\mu}\partial_{\mu}\psi-\frac{1}{2}\left[\langle W_{ij}\rangle \langle\xi\rangle^{(k_i+k_j)/2} \psi^{i}\psi^{j}\right.\nn\\
&&+\frac{2k_l}{\sqrt{h}\Lambda}\phi_1\psi_l\sigma^{\mu}\partial_{\mu}\overline{\psi}_l
+\langle W_{ij1}\rangle \frac{\langle\xi\rangle^{(k_i+k_j+2)/2}}{\sqrt{h}}\phi_1 \psi^{i}\psi^{j}\nn\\
&&\left.+ \langle W_{ijl}\rangle \langle\xi\rangle^{(k_i+k_j)/2}\phi_l \psi^{i}\psi^{j}+h.c\right]+\ldots.\,,
\end{eqnarray}
where only terms linear in $\phi_i$ are shown and $\psi$ are now restricted to lepton fields. We can use the equations
of motion to eliminate derivative interactions from the above Lagrangian. We get:
\begin{eqnarray}
\nonumber
\mathscr{L}&=&i\overline{\psi}\bar{\sigma}^{\mu}\partial_{\mu}\psi-\frac{1}{2}\left[\langle W_{ij}\rangle \langle\xi\rangle^{(k_i+k_j)/2} \psi^{i}\psi^{j}\right.\\
\label{eq:Lag_full}&&+\left[\langle W_{ij1}\rangle \langle\xi\rangle\Lambda-i(k_i+k_j)\langle W_{ij}\rangle\right]\frac{1}{\sqrt{h}\Lambda}\langle\xi\rangle^{(k_i+k_j)/2} \phi_1\psi^{i}\psi^{j}\nn\\
&&\left.+\langle W_{ijl}\rangle\langle\xi\rangle^{(k_i+k_j)/2} \phi_l\psi^{i}\psi^{j}+h.c.\right]+\ldots\,.
\end{eqnarray}
In addition, we have canonical kinetic terms for scalar fields and a generic scalar mass term.

In a complete generic setup, the scalar fields responsible for flavour symmetry breaking are both the modulus
and the flavons. We start by considering minimal models where flavons are absent and, besides Lagrangian parameters, lepton masses depend only on the modulus VEV $\langle\phi_1\rangle=\langle\Lambda\tau\rangle$.
If the only field responsible for flavour symmetry breaking is the modulus $\phi_1$, the superpotential for the charged lepton and the neutrino masses, possibly after after seesaw, can be written as:
\begin{equation}
W=E^{c}_iY^{e}_{ij}(\phi_1)L_jH_d+\frac{1}{2\Lambda_L}L_iY^{\nu}_{ij}(\phi_1)L_jH_uH_u\,,
\end{equation}
where both $Y^{e}_{ij}(\phi_1)$ and $Y^{\nu}_{ij}(\phi_1)$ are combinations of modular forms.
We can decompose the complex modulus $\phi_1$ into real and imaginary part
\begin{equation}
\phi_1=\frac{1}{\sqrt{2}} (u_1+iv_1)~,
\end{equation}
We assume a generic mixing between $u$ and $v$, due to some underlying mechanism. Their mass matrix is a general symmetric real matrix with eigenvalues $M_{u,v}$, diagonalized through the orthogonal transformation:
\be
u_1\to \cos\theta~ u_1+ \sin\theta~ v_1~,~~~v_1\to -\sin\theta~ u_1+ \cos\theta~ v_1~~~,
\ee
corresponding to $\phi_1\to e^{-i\theta}\phi_1$.
Comparing with the general formalism, we find:
\begin{eqnarray}
\label{modn1}
\nonumber&&\mathcal{Z}^{e}_u=-\frac{1}{\sqrt{2}}e^{-i\theta}T^{e}_{11},~~~~~\mathcal{Z}^{e}_v=-\frac{i}{\sqrt{2}}e^{-i\theta}T^{e}_{11}\,,\\
&&\mathcal{Z}^{\nu}_u=-\frac{1}{\sqrt{2}}e^{-i\theta}T^{\nu},~~~~~\mathcal{Z}^{\nu}_v=-\frac{i}{\sqrt{2}}e^{-i\theta}T^{\nu}~~~,
\end{eqnarray}
where
\be
\label{modn2}
T^{e}=U^{T}_{E^c}X^{e}U_{E}~~~,~~~~~T^{\nu}=U^{T}_{E}X^{\nu}U_{E}~~~.
\ee
The matrices $X^{e}$ and $X^{\nu}$ are defined as:
\begin{eqnarray}
\label{modn3}
\nonumber
X^{e}_{ij}&=&\left[i(k^{E^c}_i+k^{L}_j)Y^{e}_{ij}(\langle\phi_1\rangle)-
Y^{e}_{1ij}(\langle\phi_1\rangle) \langle\xi\rangle\Lambda\right]\frac{1}{\sqrt{h}\Lambda}\langle\xi\rangle^{(k^{E^c}_i+k^L_j)/2} v_d~~~,\nn\\
X^{\nu}_{ij}&=&\left[i(k^{L}_i+k^{L}_j)Y^{\nu}_{ij}(\langle\phi_1\rangle)-
Y^{\nu}_{1ij}(\langle\phi_1\rangle) \langle\xi\rangle\Lambda\right]\frac{1}{\sqrt{h}\Lambda}\langle\xi\rangle^{(k^{L}_i+k^L_j)/2} \frac{v^2_u}{\Lambda_L}~~~
\end{eqnarray}
with $Y^{e}_{1ij}=\frac{\partial Y^{e}_{ij}}{\partial\phi_1}$ and $Y^{\nu}_{1ij}=\frac{\partial Y^{\nu}_{ij}}{\partial\phi_1}$.
The unitary transformations $U_{E}$, $U_{E^c}$ diagonalize the charged lepton mass matrix $M^e$, while
$U_{E}$ also acts on neutrinos:
\bea
U^{T}_{E^c}M^eU_{E}&=&\text{diag}(m^e_1,m^e_2, m^e_3)~~~,~~~~~M^e_{ij}=Y^{e}_{ij}(\langle\phi_1\rangle)\langle\xi\rangle^{(k^{E^c}_i+k^{L}_j)/2}v_d~~~,\nn\\
U^T_E M^\nu U_E&=&m^\nu~~~,~~~~~~~~~~~~~~~~~~~~~~~M^\nu_{ij}=Y^{\nu}_{ij}(\langle\phi_1\rangle)\langle\xi\rangle^{(k^{L}_i+k^{L}_j)/2}\frac{v^2_u}{\Lambda_L}~~~.
\eea
In this basis, the neutrino mass matrix is diagonalized by the physical lepton mixing matrix:
$U^{T}_{PMNS}m^\nu U_{PMNS}=\text{diag}(m^\nu_1,m^\nu_2, m^\nu_3)$.
The exchange of $u_1$ and $v_1$ leads to deviations from the Newton law. If the $u$ exchange is dominant, we have
\begin{equation}
\tilde{\alpha}_u=\frac{[{\tt Re}{\cal Z}^e_u]^2}{4\pi G u^2}=\frac{[{\tt Re}(e^{-i\theta}T^{e}_{11})]^2}{8\pi G u^2}\,,
\label{alphau}
\end{equation}
otherwise
\begin{equation}
\tilde{\alpha}_v=\frac{[{\tt Re}{\cal Z}^e_v]^2}{4\pi G u^2}=\frac{[{\tt Im}(e^{-i\theta}T^{e}_{11})]^2}{8\pi G u^2}\,.
\label{alphav}
\end{equation}
The correction to the light neutrino mass matrix is given by
\begin{eqnarray}
\nonumber\delta m_\nu(0)&=&-n_e^0 \left[\frac{{\tt Re}({\cal Z}^e_u)}{M_u^2}F(M_u R) {\cal Z}^\nu_u+\frac{{\tt Re}({\cal Z}^e_v)}{M_v^2}F(M_v R) {\cal Z}^\nu_v\right]\\
&=&-\frac{n_e^0}{2} \left[{\tt Re}(e^{-i\theta}T^{e}_{11})\frac{F(M_u R)}{M_u^2}-i{\tt Im}(e^{-i\theta}T^{e}_{11})\frac{F(M_v R)}{M_v^2}\right]e^{-i\theta}T^{\nu}~~~.
\label{modshift}
\end{eqnarray}
\subsection{\label{subsec:modular_flavon}Modular Invariant models with flavons}
We extend here the previous results to the case where also flavons with vanishing weight are present. We assume that charged lepton Yukawa couplings depend on a set of flavons $\phi$, while neutrino Yukawa couplings only depend on the modulus $\phi_1$:
\begin{equation}
W=E^{c}_iY^{e}_{ij}(\phi_l)L_jH_d+\frac{1}{2\Lambda_L}L_iY^{\nu}_{ij}(\phi_1)L_jH_uH_u~~~~~~~~~(l\ne1)~~~.
\end{equation}
Decomposing the scalar fields in real and imaginary components:
\begin{equation}
\phi_l=\frac{1}{\sqrt{2}} (u_l+iv_l)~~~~~~~~(l=1,2,...),
\end{equation}
a generic orthogonal transformation acting on the basis $(u_l,v_l)$ is needed to diagonalize the scalar mass matrix\footnote{Orthogonality requires the relations $\Omega^{(uu)}\Omega^{(uu)T}+\Omega^{(uv)}\Omega^{(uv)T}=\Omega^{(vu)}\Omega^{(vu)T}+\Omega^{(vv)}\Omega^{(vv)^T}=\mathbb{1}$ and $\Omega^{(uu)}\Omega^{(vu)T}+\Omega^{(uv)}\Omega^{(vv)T}=0$.}:
\be
u_i\to\Omega^{(uu)}_{ij} u_j+\Omega^{(uv)}_{ij} v_j~~~~~~~~~~~~v_i\to\Omega^{(vu)}_{ij} u_j+\Omega^{(vv)}_{ij} v_j~~~.
\ee
we have:
\bea
\label{mif}
\mathcal{Z}^{e}_{u_n}&=&\frac{1}{\sqrt{2}}(T^{e}_m)_{11}(\Omega^{uu}+i\Omega^{vu})_{mn}~~~~~~~~~
\mathcal{Z}^{e}_{v_n}=\frac{1}{\sqrt{2}}(T^{e}_m)_{11}(\Omega^{uv}+i\Omega^{vv})_{mn}\nn\\
\mathcal{Z}^{\nu}_{u_n}&=&\frac{1}{\sqrt{2}} T^{\nu}_{1} (\Omega^{uu}+i\Omega^{vu})_{1n}~~~~~~~~~~~~~~~
\mathcal{Z}^{\nu}_{v_n}=\frac{1}{\sqrt{2}} T^{\nu}_{1} (\Omega^{uv}+i\Omega^{vv})_{1n}\nn\\
\eea
where
\be
T^{e}_m=U^{T}_{E^c}X^{e}_mU_{E}~~~,~~~~~T^{\nu}_{1}=U^{T}_{E}X^{\nu}_{1}U_{E}~~~.
\ee
The matrices $X^{e}_m$ and $X^{\nu}_{1}$ are defined as:
\begin{eqnarray}
\nonumber
(X^{e}_1)_{ij}&=&\frac{i}{\sqrt{h}\Lambda}(k^{E^c}_i+k^{L}_j)Y^{e}_{ij}(\langle\phi_l\rangle)\langle\xi\rangle^{(k^{E^c}_i+k^L_j)/2} v_d~~~,\nn\\
(X^{e}_l)_{ij}&=&
-Y^{e}_{lij}(\langle\phi_l\rangle)\langle\xi\rangle^{(k^{E^c}_i+k^L_j)/2} v_d~~~~~~~~~~~~(l>1)~~~,\nn\\
(X^{\nu}_1)_{ij}&=&\left[i(k^{L}_i+k^{L}_j)Y^{\nu}_{ij}(\langle\phi_1\rangle)-
Y^{\nu}_{1ij}(\langle\phi_1\rangle) \langle\xi\rangle\Lambda\right]\frac{1}{\sqrt{h}\Lambda}\langle\xi\rangle^{(k^{L}_i+k^L_j)/2} \frac{v^2_u}{\Lambda_L}~~~
\end{eqnarray}
with $Y^{e}_{lij}=\frac{\partial Y^{e}_{ij}}{\partial\phi_l}$ and $Y^{\nu}_{1ij}=\frac{\partial Y^{\nu}_{ij}}{\partial\phi_1}$.
As before, the unitary transformations $U_{E}$, $U_{E^c}$ diagonalize the charged lepton mass matrix $M^e$, while
$U_{E}$ also acts on neutrinos:
\bea
U^{T}_{E^c}M^eU_{E}&=&\text{diag}(m^e_1,m^e_2, m^e_3)~~~,~~~~~M^e_{ij}=Y^{e}_{ij}(\langle\phi_l\rangle)\langle\xi\rangle^{(k^{E^c}_i+k^{L}_j)/2}v_d~~~,\nn\\
U^T_E M^\nu U_E&=&m^\nu~~~,~~~~~~~~~~~~~~~~~~~~~~~M^\nu_{ij}=Y^{\nu}_{ij}(\langle\phi_1\rangle)\langle\xi\rangle^{(k^{L}_i+k^{L}_j)/2}\frac{v^2_u}{\Lambda_L}~~~.
\eea
Knowledge of the couplings in eq. (\ref{mif}) allows to estimate the shift in the neutrino mass matrix due to a region with non-vanishing electron number density, along the same lines described in the previous Section.

\section{A case study}
\label{S6}
We apply the previous results to an explicit modular invariant model of lepton masses \cite{Feruglio:2017spp}, that has been shown \cite{Ding:2019zxk} to successfully
reproduce the observed masses and mixing angles~\footnote{In ref. \cite{Ding:2019zxk} the model is labelled as ${\cal D}_{10}$.}. The model is realized at level $N=3$.
Representations and weights of the supermultiplets are listed in table \ref{samplemodel}. Neutrinos get their masses via the type I see-saw mechanism.

\begin{table}[h!]
\centering
\begin{tabular}{|c|c|c|c|c|c|c||c|}  \hline\hline
 & $L$ & $E^{c}_1$ & $E^{c}_2$ & $E^{c}_3$ & $N^{c}$ & $H_{u,d}$ & $Y_{\mathbf{3}}(\tau)$\\ \hline
$SU(2)_{L}\times U(1)_{Y}$ & $(2, -1/2)$ & $(1, 1)$ & $(1, 1)$ & $(1, 1)$ & $(1, 0)$ & $(2,\pm 1/2)$ & $(1,0)$ \\ \hline
$\Gamma_3$ & $\mathbf{3}$ & $\mathbf{1}$ & $\mathbf{1}''$ & $\mathbf{1}'$ & $\mathbf{3}$ & $\mathbf{1}$ & $\mathbf{3}$ \\ \hline
$k_{I}$ & $1$ & $1$ & $1$ & $1$ & $1$ & $0$ & $+2$\\ \hline\hline
\end{tabular}
\caption{\label{samplemodel}
Representations and modular weights of matter superfields in the benchmark model of Ref.~\cite{Feruglio:2017spp}. Also shown are the level-3 weight-2 modular forms $Y_{\mathbf{3}}(\tau)$.}
\end{table}
In a standard notation, see table \ref{samplemodel}, the superpotential for the charged lepton sector is given by:
\be
{W}_e
=\alpha E^c_1(LY^{(2)}_{\mathbf{3}})_{\mathbf{1}}H_d
+\beta E^c_2(LY^{(2)}_{\mathbf{3}})_{\mathbf{1}'}H_d + \gamma E^c_3(LY^{(2)}_{\mathbf{3}})_{\mathbf{1}''}H_d~~~,
\label{we}
\ee
where $Y_{\mathbf{3}}(\tau)$ denote the irreducible triplet of level-3 weight-2 modular forms.
The charged lepton mass matrix reads:
\begin{equation}
\label{eq:Me} m_e =\begin{pmatrix}
 \alpha Y_1(\tau) & \alpha Y_3(\tau) & \alpha Y_2(\tau) \\
\beta Y_2(\tau) & \beta Y_1(\tau)  & \beta Y_3(\tau) \\
 \gamma Y_3(\tau) & \gamma Y_2(\tau) & \gamma Y_1(\tau)
 \end{pmatrix} v_d~~~.
\end{equation}
The superpotential relevant to neutrino masses is:
\be
{W}_\nu
=g_1((N^c\,L)_{\mathbf{3}_S}Y^{(2)}_{\mathbf{3}})_\mathbf{1}H_u+g_2((N^c\,L)_{\mathbf{3}_A}Y^{(2)}_{\mathbf{3}})_\mathbf{1}H_u
+\frac{1}{2}\Lambda_L (\left(N^c N^c\right)_\mathbf{3_S}Y)_\mathbf{1}~~~.
\label{wnu}
\ee
The Dirac neutrino mass matrix $m_D$ and heavy Majorana neutrino mass matrix $m_N$ take the following form
\begin{eqnarray}
\nonumber&&\qquad\quad~~~ m_N = \begin{pmatrix}
2Y_1(\tau) ~&~ -Y_3(\tau) ~&~ -Y_2(\tau) \\
 -Y_3(\tau) ~&~ 2Y_2(\tau)  ~&~ -Y_1(\tau)  \\
 -Y_2(\tau) ~&~ -Y_1(\tau) ~&~2Y_3(\tau)
\end{pmatrix}\Lambda_L\,,\\
\label{mnuss}&&
m_D =\begin{pmatrix}
2g_1Y_1(\tau)        ~&~  (-g_1+g_2)Y_3(\tau) ~&~ (-g_1-g_2)Y_2(\tau) \\
(-g_1-g_2)Y_3(\tau)  ~&~     2g_1Y_2(\tau)    ~&~ (-g_1+g_2)Y_1(\tau)  \\
 (-g_1+g_2)Y_2(\tau) ~&~ (-g_1-g_2)Y_1(\tau)  ~&~ 2g_1Y_3(\tau)
\end{pmatrix}v_{u}\,.
\end{eqnarray}
The light neutrino mass matrix is $m_\nu=-m_D^T (m_N)^{-1} m_D$.
Charged lepton masses can be reproduced by adjusting the parameters $\alpha$, $\beta$ and $\gamma$,
while neutrino masses and the lepton mass matrix $U_{PMNS}$ depend also on additional five parameters: one overall scale, the complex combination
$g_2/g_1$ and the $\tau$ VEV. An excellent fit \cite{Ding:2019zxk} to neutrino masses and mixing angles is obtained by the
choice \footnote{These values, updating those in ref. \cite{Ding:2019zxk}, are obtained from the latest global fit of NuFIT v4.1 \cite{Esteban:2018azc,nufit}. For other global fits, see \cite{Capozzi:2020qhw}.}:
\bea
\label{modpar}
{\tt Re}(\tau)&=&0.476~~,~~~~~{\tt Im}(\tau)=1.299~~\,,\nn\\
|g_2/g_1|&=&1.210 ~~,~~~~~{\tt arg}(g_2/g_1)=4.752~~,~~~~~
\frac{|g_1|^2 v_u^2}{\Lambda_L}=0.020~{\rm eV}\,,\\
\alpha v_d&=&102.253 {\rm MeV}~,~~~\beta v_d = 1753.220 {\rm MeV}~,~~~
 \gamma v_d = 0.501 {\rm MeV}~~,\nn
\eea
for normally ordered neutrino mass spectrum. This model can also accommodate inverted ordering neutrino mass spectrum which is disfavored by the present data, and we shall not discuss this case in the present work. Neutrino masses and mixing parameters at the best fit point, eq. (\ref{modpar}) are shown in Table \ref{fitresults}.

\begin{table}[h!]
\centering
\begin{tabular}{|c|c|c|c|c|c|}
\hline\hline
$\sin^2\theta_{12}$&$\sin^2\theta_{13}$&$\sin^2\theta_{23}$&$\delta_{CP}/\pi$&$\alpha_{21}/\pi$&$\alpha_{31}/\pi$\\
\hline
$0.3105$&$0.0224 $&$0.5631$&$1.4821$&$1.3089$&$1.4541$\\
\hline
$m_1({\rm eV})$&$m_2({\rm eV})$&$m_3({\rm eV})$&$\Delta m^2_{21}({\rm eV})$&$\Delta m^2_{31}({\rm eV})$&$|m_{ee}|({\rm eV})$\\
\hline
$0.0409$&$0.0418$&$0.0647$&$7.39\times 10^{-5}$&$2.522\times 10^{-3}$&$0.0223$\\
\hline\hline
\end{tabular}
\caption{\label{fitresults} Values of neutrino masses and mixing parameters at the best fit point, eq. (\ref{modpar}), obtained from the latest global fit of NuFIT v4.1 \cite{Esteban:2018azc,nufit}.}
\end{table}
We evaluate the shift $\delta m_\nu(0)$ in eq. (\ref{modshift}) at the center of the sun,
taking $R=6.955\times 10^5$ Km and $n_e(r)$ from ref. \cite{Bahcall:2004pz,Bahcall_web},
by assuming the $v_1$ component
of the modulus sufficiently heavy to escape existing bounds on $v$ couplings.
The $u_1$ exchange is dominant, eq. (\ref{alphau}) applies and $\delta m_\nu(0)$ is saturated by
the first contribution in eq. (\ref{modshift}):
\be
\label{shiftmod1}
\delta m_\nu(0)=-\frac{n_e^0}{2} {\tt Re}(e^{-i\theta}T^{e}_{11})\frac{F(M_u R)}{M_u^2}e^{-i\theta}T^{\nu}~~~.
\ee
The other possible case, when $u_1$ decouples and $v_1$
dominates both the long range force and the mass shift $\delta m_\nu(0)$, is obtained from the former case through the parameter redefinition $\theta\rightarrow\pi/2+\theta$, $M_{u}\rightarrow M_v$. Thus, without loss of generality, we can focus on the $u_1$-dominated scenario. Expressing ${\tt Re}(e^{-i\theta}T^{e}_{11})$ in terms of $\tilde{\alpha}$, we find the neutrino mass shift $\delta m_\nu(0)$ is given by
\begin{equation}
\label{pm}
\delta m_\nu(0)=\pm n_e^0\sqrt{2\pi G u^2\tilde{\alpha}_u}\frac{F(M_u R)}{M_u^2}e^{-i\theta}T^{\nu}\,,
\end{equation}
where the ``+'' and ``$-$'' signs correspond to $\theta-{\tt arg}(T^e_{11})=\arccos(\sqrt{8\pi G u^2\tilde{\alpha}_u}/|T^e_{11}|)$ and  $\theta-{\tt arg}(T^e_{11})=\pi-\arccos(\sqrt{8\pi G u^2\tilde{\alpha}_u}/|T^e_{11}|)$
respectively. From the eqs. (\ref{modn1}-\ref{modn3}), we see that $T^{\nu}$ is of order $m_{\nu}/\Lambda$. To obtain an observable effect while keeping $|{\tt Re}({\cal Z}^e_u)|=|{\tt Re}(e^{-i\theta}T^{e}_{11})|/\sqrt{2}$ close to $10^{-25}$, we need $\Lambda\approx{10^9}$ GeV and $M_u\leq10^{-16}$ eV. The scalar neutrino coupling is of order $m_{\nu}/\Lambda\approx 10^{-20}$, safely below the current limits.
At the same time, since $|T^e_{11}|\approx m_e/\Lambda\approx 10^{-12}$, the phase difference
$|\theta-{\tt arg}(T^e_{11})|$ should be kept very close to $\pi/2$.
It turns out that in the model under study, ${\tt arg}(T^e_{11})\approx\frac{\pi}{2}$ at the best fit point, such that the angle $\theta$ is require to be around 0 or $\pi$. This means there is almost no mixing between $u_1$ and $v_1$.

\begin{figure}[h!]
\centering
\includegraphics[width=0.48\textwidth]{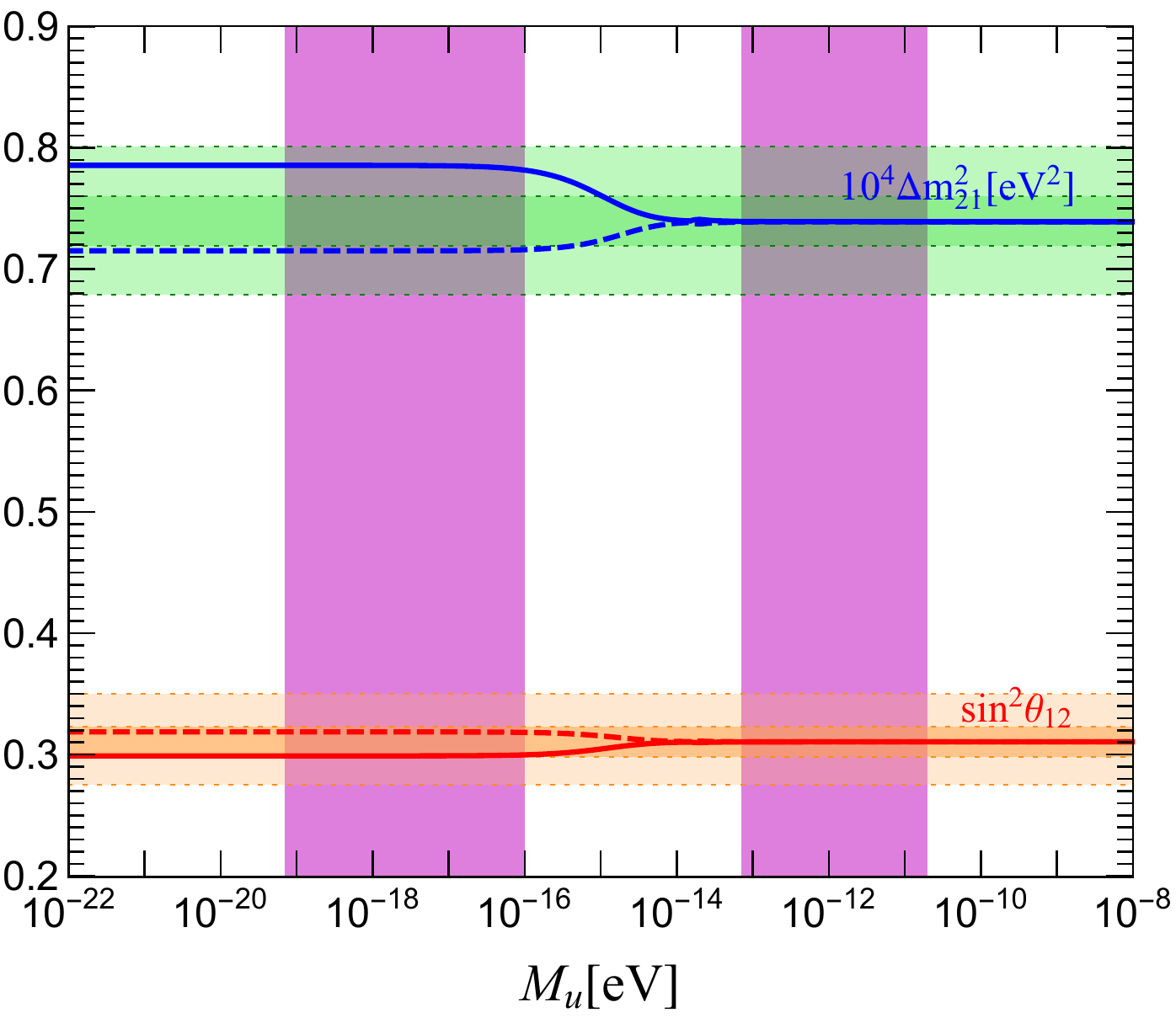}\quad
\includegraphics[width=0.48\textwidth]{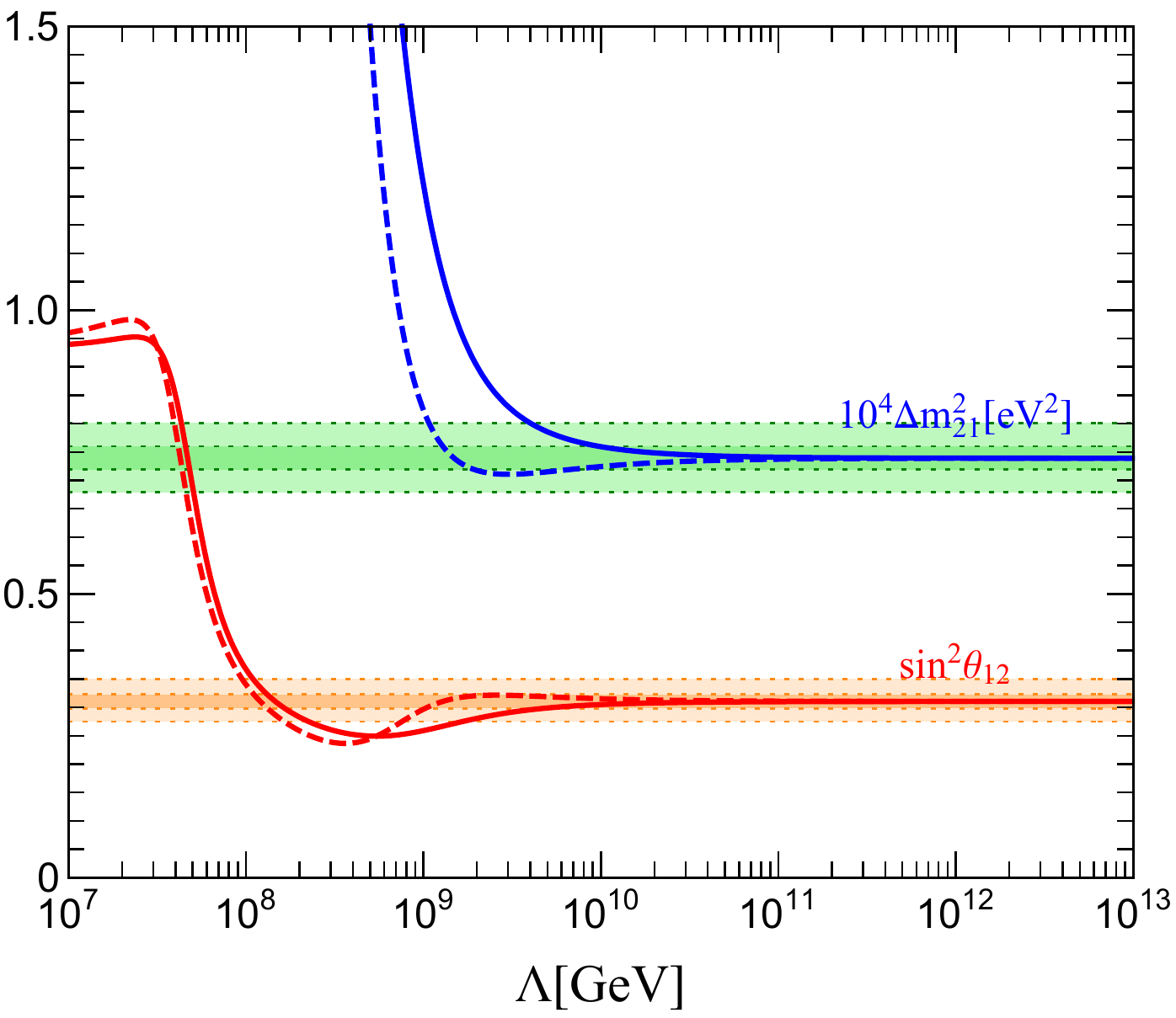}
\caption{$\Delta m^2_{21}$ and $\sin^2\theta_{12}$ versus $M_u$ for $\Lambda=5\times10^9$ GeV (left panel) and versus $\Lambda$ for $M_u=10^{-22}$ eV (right panel). We set $h=1$ and the angle $\theta$ has been tuned to suppress the scalar-electron coupling below the experimental bound. Solid(dashed) line for plus(minus) sign in eq. (\ref{pm}), respectively. The darker(lighter) green and orange bands show the present 1$\sigma$(3$\sigma$) allowed region from~\cite{Esteban:2018azc,nufit}. The combination $n_e^0 F(M_\alpha R)$ is the one of eq. (\ref{eq:FF}), with $n_e(r)$ from ref. \cite{Bahcall:2004pz,Bahcall_web}.
The vertical bands in purple are excluded from black hole superradiance.}
\label{plotmod1}
\end{figure}\vskip 0.5 cm
\noindent

Fig. \ref{plotmod1} shows the most important deviations from scalar NSI.  They affect the solar oscillation parameters $\Delta m^2_{21}$ and $\sin^2\theta_{12}$. In these plots we have fixed  $|{\tt Re}{\cal Z}^e_u|=\sqrt{4\pi G u^2\tilde\alpha}$ such that the bounds on $\tilde{\alpha}$ extracted from the tests of ISL and EP are satisfied. This is always possible by tuning the angle $\theta$. The smaller $\Lambda$, the higher the degree of the tuning required. For $\Lambda$ larger than about $10^{11}$ GeV, $\Delta m^2_{21}$ and $\sin^2\theta_{12}$ are essentially unchanged due to the smallness of the scalar-neutrino coupling and the saturation effect due to the factor $F(M R)$. For fixed and sufficiently small values of the scalar mass, such as $M_u=10^{-22}$ eV, the deviations for $\Delta m^2_{21}$ and $\sin^2\theta_{12}$ can be very large, as shown in the right panel. One of the reason for such a behavior is that the model predicts nearly degenerate $m_{1,2}$ neutrino masses with a mass difference of about 0.0009 eV, see table \ref{fitresults}. It suffices to perturb these masses in one part over one hundred to upset the prediction for $\Delta m^2_{21}$ and similarly for $\sin^2\theta_{12}$. Moreover, as we see from the right panel of fig. \ref{plotmod1}, for sufficiently small values of $\Lambda$, the neutrino mass shift $\delta m_{\nu}(0)$ dominates over the leading order neutrino mass matrix and the neutrino mixing parameters receive huge corrections.

\begin{figure}[h!]
\centering
\includegraphics[width=0.49\textwidth]{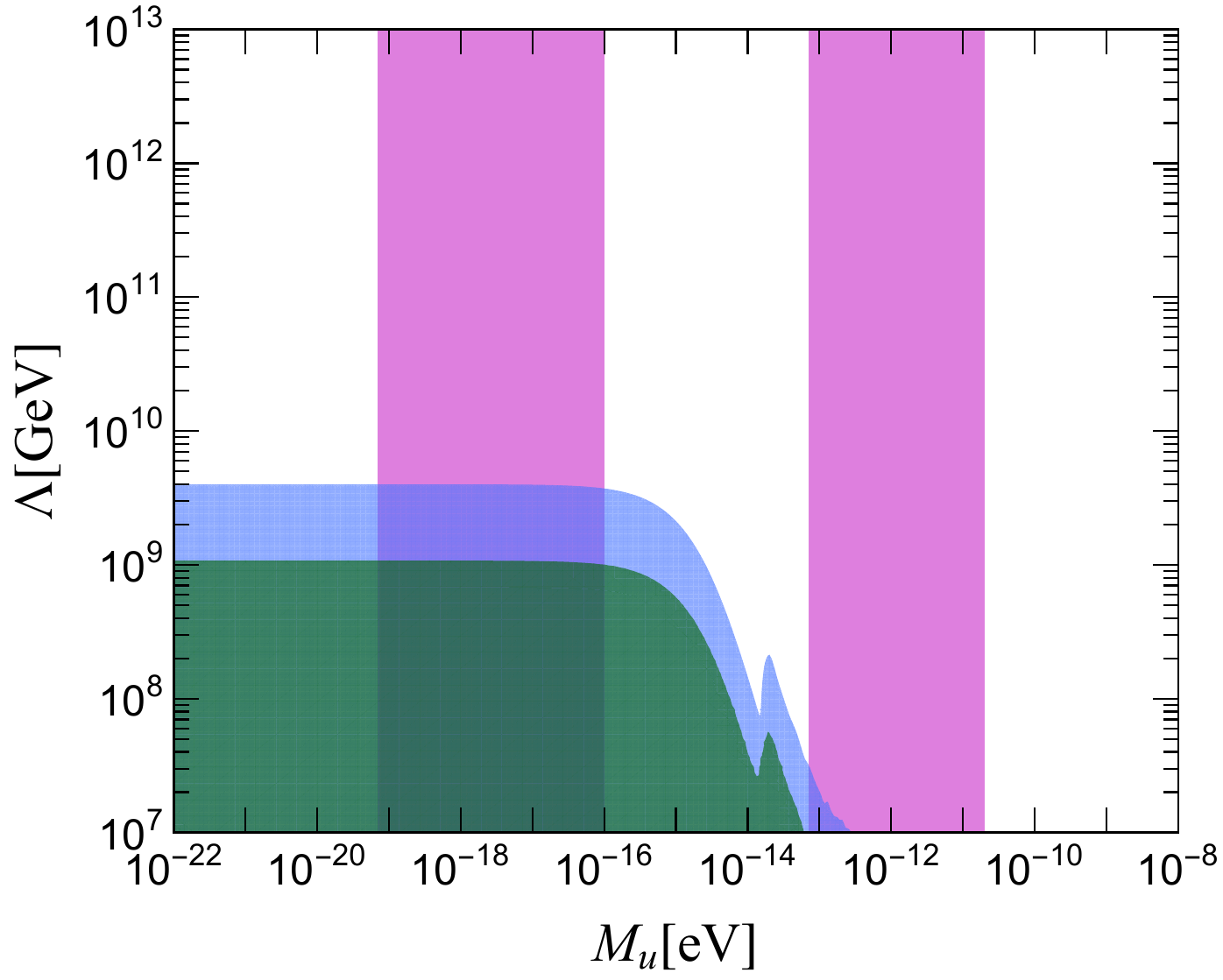}~
\includegraphics[width=0.49\textwidth]{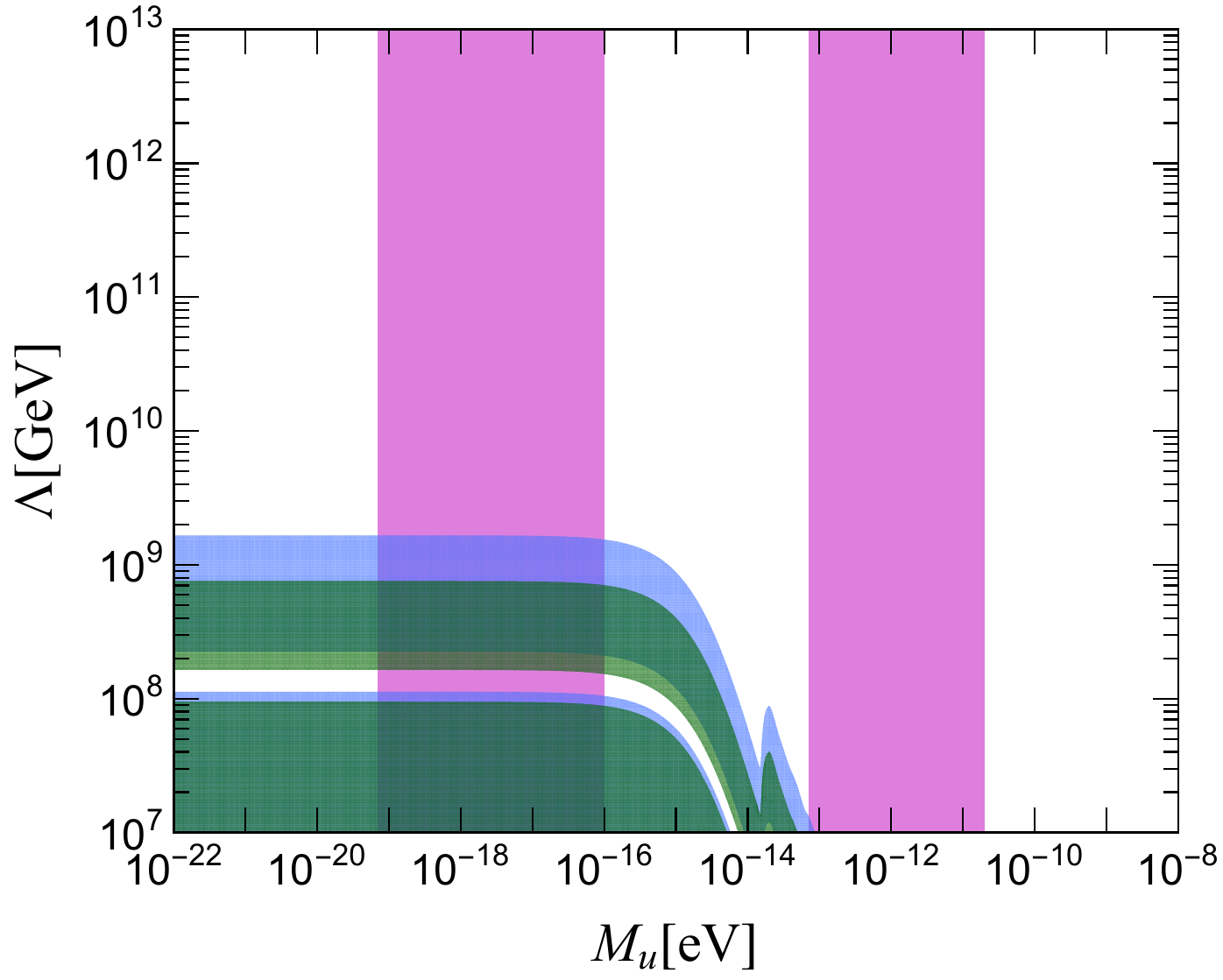}
\caption{Exclusion regions in the plane $(M_u,\Lambda)$ from the data of $\Delta m^2_{21}$ (left panel) and $\sin^2\theta_{12}$ (right panel), see text for explanation. The blue and green areas are for the plus and minus signs in eq. (\ref{pm}) respectively. The vertical bands in purple are excluded from black hole superradiance.}
\label{fig:plotmod2}
\end{figure}

In fig.~\ref{fig:plotmod2} we estimate the region of the parameter space already excluded by the experimental data of $\Delta m^2_{21}$ and $\sin^2\theta_{12}$ at $3\sigma$ level~\cite{Esteban:2018azc,nufit}. An accurate determination of such a region would require a full simulation of neutrino oscillations in the sun, with neutrino masses and mixing angles varying along the sun profile, which goes beyond the scope of this work. In our estimate, we compare the values of  $\Delta m^2_{21}$ and $\sin^2\theta_{12}$ evaluated at the center of the sun with the results of the most recent global fit and we declare excluded the parameters leading to a deviations larger than 3 $\sigma$. The angle $\theta$ is tuned to keep the scalar-electron coupling at the largest value allowed by the current bounds, while the parameter $h$ is fixed to 1.
Within this simple-minded approach, we see that a sizable portion of the parameter space of the model is already excluded by the present data. We see that $\Delta m^2_{21}$ is more effective than $\sin^2\theta_{12}$ to constrain the model.

\vskip 0.5 cm
\noindent

The above results apply to a specific model, but probably some lessons can be extended to the full class of modular invariant models. If the only flavon is the modulus, models in this class have the same number of extra parameters, beyond those required to fit oscillation data. From dimensional analysis we expect a similar dependence of the coupling ${\cal}Z^e$ and ${\cal}Z^\nu$ on the lepton masses and the cutoff scale $\Lambda$. Thus we foresee qualitatively similar effects from scalar NSI. Moreover, since in this class of models $m_1$ and $m_2$ are typically very close, independently on the type of mass hierarchy, we expect that large deviations in the $(\Delta m^2_{21},\sin^2\theta_{12})$ sector due to scalar NSI are possible.
If we consider also models falling in the framework discussed in section \ref{subsec:modular_flavon}, where charged leptons and neutrinos
get their masses from two different scalar sectors, characterized by different flavour scales, it might be possible to alleviate the fine tuning required to adequately suppress the scalar-electron coupling.

\section{Discussion}
\label{S7}
New scalar particles are naturally expected in most of SM extensions attempting to explain the flavour puzzle. Common to these extensions is the concept that the observed Yukawa couplings originate from the VEVs of field-dependent quantities. The new scalar particles mediate new, so far undetected, interactions among the SM fermions, described by higher-dimensional operators depleted by the scale of flavour dynamics.
To avoid new detectable sources of flavour changing neutral currents, such scale is often assumed to be much larger than the electroweak scale. Nevertheless at least a portion of the parameter space of the above scenario can still be tested in present experiments. Indeed, in the lepton sector scalar exchange gives rise to scalar NSI, resulting is a shift of the neutrino mass matrix when neutrinos propagate in a medium with non-vanishing electron number density. If sufficiently large, this shift could alter the standard picture of neutrino oscillations and lead to an observable effect,
for instance in solar neutrino oscillations.

In the center of a spherical region of radius $R$ with uniform electron number density $n_e^0$ neutrinos experience a mass shift
\be
\label{summary}
\delta m_\nu(0)=-n_e^0   \frac{{\tt Re}({\cal Z}^e){\cal Z}^\nu}{M^2(R)} ~~~.
\ee
where ${\cal Z}^e$ and ${\cal Z}^\nu$ are the couplings of the scalar field to electrons and neutrinos and $M(R)$ is the effective scalar mass, approximately equal to ${\tt max}(M,\hbar/(R c))$. The important point that $M(R)$ is not simply the mass of the scalar particle, but a scale bounded by the inverse size of the region where electrons are concentrated,
has been recently highlighted in ref. \cite{Smirnov:2019cae}.
To produce a shift of few meV in a region with an electron number density close to the one in the sun,
an effective coupling ${\tt Re}({\cal Z}^e){\cal Z}^\nu/M(R)^2\approx 10^4$ GeV$^{-2}$ is required, more than eight orders of magnitude larger than the Fermi constant. To maximize the effect we are led to consider extremely light scalar mediators. In the sun the smallest value of $M(R)$ is approximately $\hbar/(R c)=3\times 10^{-16}$ eV, realized for any scalar mass $M$ smaller than or equal to $\hbar/(R c)$. Scalar masses in the window $(10^{-19}\div 10^{-16})$ eV are excluded by experiments looking for super-massive black-hole superradiance.

Even in presence of the huge enhancement provided by such small mediator mass, the size of the effect is severely bounded by the existing limits on scalar-electron and scalar-neutrino couplings. For tiny scalar masses, the limits on scalar-electron couplings are dominated by the negative results
of the search for new long-range forces. In the present work we included the results of the MICROSCOPE experiment that, for scalar masses below about $1.5\times10^{-14}$ eV, has set the strongest bound on the scalar-electron Yukawa coupling: $|{\tt Re}{\cal Z}^e|<2.4\times 10^{-25}$. Scalar-neutrino couplings are constrained by cosmological data. In the history of the universe, scalars with masses of the order $3\times 10^{-16}$ eV or below can be treated as massless, their interaction rate is proportional to the universe temperature $T$ and become relevant only after neutrino decoupling. From CMB data the bounds $|{\cal Z}^\nu_{ii}|<1.2\times 10^{-7}$ and
$|{\cal Z}^\nu_{ij}|<2.3\times 10^{-11} (0.05~{\rm eV}/m)^2$ [$m={\tt max}(m_i,m_j)~(i\ne j)$] follow. They are stringent, but much less than the one adopted by ref. \cite{Smirnov:2019cae}: $|{\cal Z}^\nu|^2/M^2<(3 ~{\rm MeV})^{-2}$ or $|{\cal Z}^\nu|<10^{-22} \times M({\rm eV})/(3\times 10^{-16})$. The different set of bounds adopted here is at the origin of new numerical results and different conclusions. If no further restrictions other than the experimental ones apply to the relevant parameters, from eq. (\ref{summary}) we see that there is a considerable region in parameter
space where the shift of neutrino masses due to scalar NSI is observable in solar neutrino oscillations.

Although the previous conclusion is rather encouraging, it is highly desirable to verify whether such region of parameters is favoured or not in physically motivated extensions of the SM. For this reason, in this paper we have analyzed scalar NSI in a specific class of models, aiming at the description of lepton masses and mixing parameters within the framework of broken flavour symmetries. In this context the scalar particles are nothing but the flavons, or the modulus in modular invariant models. A detectable effect from scalar NSI would allow us to access the otherwise elusive dynamics of the flavon sector.

We have proceeded under the working assumption that in these models the mass of one of the flavons or modulus can be as small as $3\times 10^{-16}$ eV. Such a fantastic suppression
compared to other known scales might pose a new hierarchy problem, but is not contradicted by experiments.
Concerning the couplings of such a light scalar, in the class of models investigated here the functional dependence of Yukawa couplings is constrained by the flavour symmetry
and the relevant parameters ${\cal Z}^e$ and ${\cal Z}^\nu$ cannot take arbitrary values. Very roughly,
at the level of order of magnitudes, in the presence of a single scalar field $\varphi$ acquiring the VEV $\varphi^0$ we have:
\bea
{\cal Z}^e&\approx& \frac{m_e}{\varphi^0}\approx 5\times 10^{-20}\left(\frac{10^{16}}{\varphi^0({\rm GeV})}\right)~~~,\nn\\
{\cal Z}^\nu&\approx& \frac{m_\nu}{\varphi^0}\approx 5\times 10^{-27} \left(\frac{m_\nu({\rm eV})}{0.05}\right)\left(\frac{10^{16}}{\varphi^0({\rm GeV})}\right)~~~.
\label{cou}
\eea
A rough estimate of the expected shift, gives the result:
\be
\frac{\delta m_\nu(0)}{m_\nu}\approx -0.006 \left(\frac{n_e^0({\rm eV}^3)}{10^{11}}\right)\left(\frac{3\times 10^{-16}}{M({\rm eV})}\right)^2\left(\frac{10^{16}}{\varphi^0({\rm GeV})}\right)^2~~~.
\label{est}
\ee
From eqs. (\ref{cou}) we see that, for reasonable values of the scalar VEV $\varphi^0$, the electron-scalar coupling
cannot satisfy the bound set by the MICROSCOPE experiment. To verify the existence of parameters allowing an observable effect and not excluded by the present limits, we have explored more carefully specific symmetry realizations.
We have analyzed two models where scalar NSI can be potentially detected. In the first model the flavour symmetry
is abelian. The second one is modular invariant and provides an excellent fit to the observed neutrino masses
and lepton mixing angles in terms of five parameters. In both models the electron-scalar coupling is suppressed below the existing limits by a mixing angle describing the fraction of
the ultra-light scalar that couples to the electron. In the first model the neutrino-scalar coupling can be even enhanced with respect to the estimate in eq. (\ref{cou}) by the VEV of an independent scalar multiplet.
In general, the desired suppression might also be induced by an appropriate mixing between the flavon/modulus and the Higgs field. In both models observable effects are achievable in solar neutrino oscillations while respecting all experimental bounds. Modular invariant models typically predict nearly degenerate $m_{1,2}$ neutrino masses, with $m_2-m_1$ of the order of 1 meV, independently on the type of mass hierarchy, As a consequence, even small corrections to the neutrino mass matrix induced by scalar NSI, can result in sizable effect at the level of the solar oscillation parameters $(\Delta m^2_{21},\sin^2\theta_{12})$.

The major obstacle to observability is represented by the extremely small value of the scalar-electron coupling, requiring
an additional suppression factor beyond the one provided by the scalar VEV in eq. (\ref{cou}).
The ingredients of such extra suppression are present in most of the existing constructions, being related to the expected mixing in the
scalar sector. Though almost unavoidably present, such a mixing must however be accurately tuned  to provide the desired set of couplings.
The region of parameter space surviving the experimental bounds is limited, but has not yet shrank to zero.
The scalar sector of models based on flavour symmetries is often designed only to produce a suitable set of VEVs
and its dynamics is neglected in most of the cases, especially if the involved breaking scales are very large.
The detection of effects from scalar NSI would represent a major accomplishment and would open the way
to directly access the flavon dynamics. Moreover the shift of the predicted neutrino mass matrix
is closely related to the flavour symmetry pattern, thus providing additional precious information.

\section*{Acknowledgements}
We thank Marco Peloso for useful comments and Pierre Fayet for useful correspondence and suggestions on the limits applying to new long range forces and in particular for drawing the results of the MICROSCOPE experiment to our attention. This project has received support in part by the MIUR-PRIN project 2015P5SBHT 003 ``Search for the Fundamental Laws and Constituents'' and by the European Union's Horizon 2020 research and innovation programme under the Marie Sklodowska-Curie grant
agreement N$^\circ$~674896 and 690575 and by the National Natural Science Foundation of China under Grant Nos 11975224, 11835013, 11947301. The research of F.~F.~was supported in part by the INFN. F.~F.~thanks the University of Science and Technology of China (USTC) in Hefei for hospitality in July 2019, when this project started. GJD thanks the  Department of Physics and Astronomy, University of Padova for hospitality in January 2020.

\providecommand{\href}[2]{#2}\begingroup\raggedright\endgroup

\end{document}